\documentclass[prb,aps,twocolumn,preprintnumbers,amsmath,amssymb,array]{revtex4}

\usepackage{graphicx}
\usepackage{dcolumn}
\usepackage{bm}

\DeclareGraphicsExtensions{.png,.jpg,.pdf}

\begin{document}
\title{Constructing Spin Interference Devices from Nanometric Rings}
\author{Guy Cohen$^1$, Oded Hod$^2$, and Eran Rabani$^3$}

\affiliation {$^1$ School of Physics and Astronomy, The Sackler
Faculty of Exact Sciences, Tel Aviv University, Tel Aviv 69978, Israel
{{~\\}} $^2$ Department of Chemistry, Rice University, Houston, TX
77251-1892 {{~\\}} $^3$ School of Chemistry, The Sackler Faculty of
Exact Sciences, Tel Aviv University, Tel Aviv 69978, Israel}

\date{\today}

\begin{abstract}
The study of nanospintronic devices utilizing coherent transport
through molecular scale multiply-connected geometries in the presence
of moderate magnetic fields is presented.  It is shown how two types
of simple devices, spin filters and spin splitters (or Stern-Gerlach
devices) may be constructed from molecular nanometric rings utilizing
the Aharonov-Bohm effect.  The current is calculated within a single
electron approximation and within a many-body master equation approach
where charging effects are accounted for in the Coulomb Blockade
regime.  We provide rules and tools to develop and analyze efficient
spintronic devices based on nanometric interferometers.
\end{abstract}
\maketitle

\newpage 

\section{introduction}
\label{sec:intro}
In recent years a great deal of attention has been devoted to the
study of useful electronic devices utilizing the
Aharonov-Bohm~\cite{Aharonov1959} (AB) phase in multiply-connected
geometries.\cite{Webb85,Timp87,Yacoby95,Oudenaarden98,Fuhrer01,Smalley04,Ensslin05}
In particular, there has been interest in spin-sensitive
devices~\cite{Takayanagi99,Richter01} that are the single-electron
analogue of semiconductor spintronics.\cite{Malajovich2001} Most of
the research conducted in this direction has focused on {\em
mesoscopic} systems, where the AB flux quantum matches weak magnetic
fields, inter-electronic dependencies can be relatively negligible,
and the Rashba~\cite{Rashba60} and Dresselhaus~\cite{Dresselhaus55}
spin-orbit coupling or inhomogeneous magnetic
fields,\cite{Richter04a,Richter04b} provide a large and controllable
dependency of the electron Hamiltonian on directionally significant
spin
eigenvalues.\cite{Richter01,Klapwijk02,D'Amico03,Richter04c,Vasilopoulos04,Sigrist05,Vasilopoulos05,Peeters06,Perroni06}

At the {\em nanometric} scale, it has been shown that {\em
spin-independent} AB molecular interferometers may be possible at
reasonable magnetic fields when the coupling of the device to the
leads is small.\cite{Hod2004,Hod2005a,Hod2005b,Hod2006a} For such
small AB interferometers there are striking differences in the
properties of magnetic versus electric gauges, and the magnetic gate
becomes advantageous over electrical gating.\cite{Hod2005b,Hod2006b}
For example, the current in a multi-terminal molecular device can be
tuned by changing the polarity of the magnetic field utilizing its
symmetry breaking nature.\cite{Hod2005b} Another example includes
fundamental differences between magnetic and electric gates with
respect to inelastic effects.\cite{Hod2006b}

Previous studies of nanometeric molecular AB interferometers have
ignored the spin degree of freedom.  In such molecular-scale systems,
if the likes of atomic spin-orbit coupling and magnetic ions are not
prevalent,\cite{Gefen89,Gefen92} the lack of a Rashba/Dresselhaus
field and the difficulties in locally manipulating the external
magnetic field at such scales leaves only the much smaller Zeeman term
to differentiate between spins.\cite{Richter03} In addition, the small
size of such devices further complicates matters by introducing large
charging effects.\cite{Elhassid00} These should become especially
noticeable when the coupling between the device and the leads is weak
(as required for nanometeric devices operating at reasonable magnetic
fields), since electrons spend a longer time on the device.  Despite
these inconveniences, one may very well wonder whether it is possible
to develop molecular AB interferometers as single-electron
spin-devices such as a spin filter or splitter.  This is the question
that interests us here.

In this paper we develop the basic physical foundation to develop such
devices.  In Section~\ref{sec:devices} we describe the types of
devices we wish to model and define their efficiency in terms of the
spin-dependent current.  This includes a two-terminal spin filter and
a three-terminal spin splitter.  In
Section~\ref{sec:independent-electron-calculations} Zeeman splitting
is introduced into the two models developed by Hod {\em et
al.}~\cite{Hod2004,Hod2005b,Hod2006a} for independent spinless
electrons - a simple continuum one-dimensional model and a more
detailed tight-binding model.  Within the simple continuum model we
provide an exact solution for the spin-dependent conductivity for the
two- and three-terminal devices.  The role of the different model
parameters is studied and comparison between the simple continuum
model and the tight-binding results are made.  In
Section~\ref{sec:coulomb-blockade-calculation} the tight-binding
levels are used in a master equation
calculation~\cite{Beenakker91,Kinaret92,Bonet02,Hettler03,Datta2004,Elste05,Braig05,Mukamel06,Hanggi2006,Nitzan07}
to examine charging effects in the Coulomb blockade regime.  From this
basic analysis some interesting conclusions are drawn regarding the
properties and limitations of nanospintronic AB devices, and a few
potential avenues for further research are pointed out in
Section~\ref{sec:conclusions-directions-for-further}.

\section{Devices}
\label{sec:devices}
\begin{figure}[htp]
\includegraphics[width=8cm]{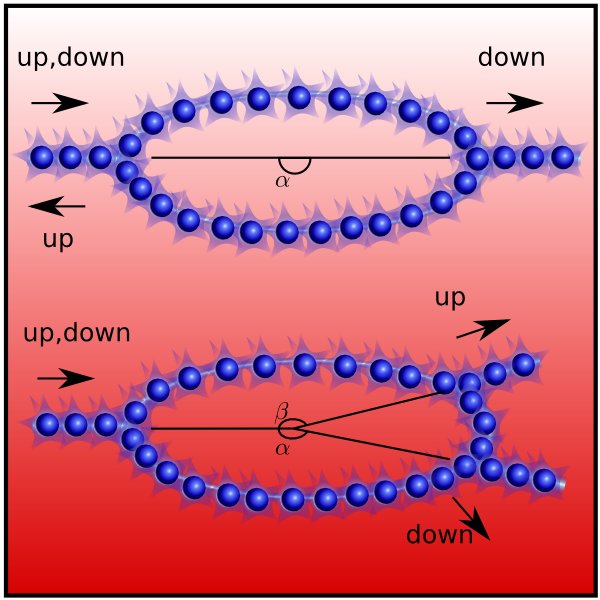}
\caption{A sketch of the two prototype devices: a molecular spin
filter (upper panel) and a molecular spin splitter (lower
panel). Up/down refers to the two spin polarized states.}
\label{fig:sketch}
\end{figure}
We will examine two types of elementary nanometric devices with and
without charging effects: a spin filter and a spin splitter, as
illustrated in Fig.~\ref{fig:sketch}.  Such devices might be built
from molecular rings such as the polycyclic aromatic
hydrocarbon,\cite{Hod2005a,Staab1978,Hajgato2004} or from atomic
corrals (with the advantage of controlled structure and electron
density).\cite{Collins1993,Manoharan00} Such corrals can be
constructed atom by atom using scanning tunneling microscopy
techniques.\cite{Ho03} A more exotic example involves 3D
nanostructures such as torus knots, where a periodicity of more than
$2\pi$ can be achieved, producing a stronger AB effect than simple
rings.  In such molecular structures, control over the structure's
geometry is possible at the atomic level (unlike the case of
mesoscopic rings).

The filter does not really require something as complex as an AB ring
for its realization.  One only requires two single-spin levels which
can, by manipulation of the magnetic field, be moved in and out of the
conduction energy window, and any atom, quantum dot, or other discrete
system can meet this requirement.  This case is treated because it
provides a simple example with only two terminals and yet may still be
useful.  The splitter is a molecular Stern-Gerlach device and is
somewhat more interesting: having three terminals, it actually
requires a device which, like an AB ring, has some sort of inherent
asymmetry in either its construction or in the spinor wavefunction.

It is useful at this point to examine the energy scales of the
problem.  The Fermi energy is the most important parameter.  For a
half-filled conduction band the Fermi wavelength is of the order of
four bond lengths $\lambda_{F}=4a$.  This gives a Fermi energy of
$\epsilon_{F}= \frac{h^2}{2m^{\star}\lambda_{F}^{2}} \approx 10
\mbox{eV}$ if we set the lowest level to zero, regardless of ring size
and assuming atomic spacing of about $a = 2 \mbox{\AA}$.  To a good
approximation, the molecular orbital energy (at zero magnetic field
$B=0$) is given by an effective mass model $\epsilon_{m} =
\frac{\hbar^2 m^2}{2m^{\star}R^2} $, where $m =0, \pm 1, \cdots$, $R$
is the radius of the ring and $m^{\star}$ is the effective mass.  This
gives a value for the Fermi quantum number of $m_{\mbox{\tiny
F}}=\frac{\pi R}{a}$.  Electronic levels on the rings near the Fermi
energy are separated by $\Delta \epsilon=\frac{\hbar^2\pi}{m^{\star}R
a}$, which reduces linearly with $R$.  For rings of the order of
several nanometers in diameter the separation between levels near the
Fermi energy is of the order of sub electron volts.  This is also the
order of magnitude of the charging energies $\epsilon_{c}$.  Other
energy scales involve the magnetic field splitting which can be as
large as the level spacing and the Zeeman splitting $2.89\cdot10^{-5}
\mbox{eV}$ per Tesla, which is only about a third of the thermal
energy at $1 \mbox{K}$.

The effectiveness of these devices can be judged not only by output
polarization, but also by unitarity: a perfect device always sends the
correct electrons into the correct lead.  More precisely, the two
devices that will be discussed are judged by the following quantities:
\begin{subequations}
\begin{eqnarray}
 & & e_{\uparrow
 filter}\equiv\frac{j_{\uparrow}}{j_{\uparrow}+j_{\downarrow}}\label{eq:1}\\
 & & e_{\downarrow
 filter}\equiv\frac{j_{\downarrow}}{j_{\uparrow}+j_{\downarrow}}\label{eq:2}\\
 & &
 e_{splitter}\equiv\frac{(j^{1}_{\uparrow}-j^{1}_{\downarrow})(j^{2}_{\downarrow}-j^{2}_{\uparrow})}{(j^{1}_{\uparrow}+j^{1}_{\downarrow})(j^{2}_{\downarrow}+j^{2}_{\uparrow})}.
\label{eq:effectivness}
\end{eqnarray}
\end{subequations}
Here $j_{\uparrow \downarrow}$ is the up/down current for the two
terminal device and $j^{1,2}_{\uparrow \downarrow}$ is the up/down
current for channel $1$ or $2$ for the three terminal device.  The
only important properties these quantities need to have for the
purpose of this work is that they should be bounded from above by $1$,
and should reach $1$ only in the case of perfect operation.  For the
two and three terminal devices both requirements are satisfied.  We
refer to them from now on as ``efficiencies''.

The problem of building a perfect device can thus be mathematically
restated as the problem of optimizing its efficiency to the desired
value, usually unity, over the space of all controllable parameters.
These include the magnetic field, the ring's radius and structure, the
gate and bias voltage (or the chemical potential) and the lead
placement and coupling.  Reasonable ranges for these parameters must
be assumed: magnetic fields of more than a few Tesla may pose a
technical limitation, as are large rings where disorder effects begin
to dominate.\cite{Aronov1987} The leads cannot be too close to one
another to avoid direct tunneling and should be coupled strongly
enough to make the environmental coupling unimportant.  Even within
these limits, the problem remains numerically formidable in all
models.  We will therefore also discuss a conceptually simpler if less
systematic way of designing perfect devices.

\section{Single Electron Picture}
\label{sec:independent-electron-calculations}
In this section we describe the physical principles required to
construct molecular spin filter and spin splitter devices in the limit
where charging effects can be ignored and the complexity of the
many-body physics can be reduced to a simplified one-electron picture.

\subsection{Basic Concepts}
To discuss the basic physical principles involved, let us first regard
a simple analytical model for the transmission through an AB
ring.\cite{Richter01,Richter04a,Richter04b,Richter04c,Peeters06} The
physics of this model for molecular conductors was recently discussed
by Hod {\em et al.} in the absence of Zeeman splitting and spin-orbit
coupling.\cite{Hod2004,Hod2005a,Hod2005b,Hod2006a} Consider a one
dimensional AB interferometer as shown in Fig.~\ref{fig:sketch}.  It
consists of a conducting ring of radius $R$ coupled to two or three
conducting wires placed in a perpendicular uniform magnetic field,
${\bf B}$.  The device is described by the Pauli Hamiltonian:
\begin{equation}
H_{s} = \frac{\mathbf{\boldsymbol{\Pi}}^{2}}{2m_{e}} + V({\bf r}) +
g\mu_{\mbox{\tiny B}}\mathbf{\boldsymbol{\sigma}}\cdot\mathbf{B},
\label{eq:hmil}
\end{equation}
where $\mathbf{\boldsymbol{\Pi}} = {\bf P}+\frac{e}{c}{\bf A}({\bf
r})$, ${\bf A}({\bf r})$ is the vector potential (${\bf B}={\bf
\nabla} \times {\bf A}$), $V({\bf r})$ is the electrostatic potential
on the ring, and $m_{e}$ is the electron mass.  The last term in
Eq.~(\ref{eq:hmil}) represents the coupling of the magnetic field to
the spin angular momentum, where $\mathbf{\boldsymbol{\sigma}}$ are
the Pauli matrices, $g$ is the gyromagnetic ratio (we take $g=2$ for
the spin), and $\mu_{\mbox{\tiny B}}=\frac{e\hbar}{2m_{e}}$ is the
electron Bohr magneton.  The lack of a Rashba field in molecular
conductors allows us to safely ignore spin-orbit coupling effects.  In
the common case where the scalar potential $V(r)$ governing the system
is periodic or nearly so, the effective mass approximation can be used
and the Pauli Hamiltonian can be reduced to:
\begin{equation}
H_{s} = \frac{\mathbf{\boldsymbol{\Pi}}^{2}}{2m^{\star}} + g
\mu_{\mbox{\tiny B}}\mathbf{\boldsymbol{\sigma}}\cdot\mathbf{B}=
H_{ring} + H_{\sigma},
\label{eq:hmil-em}
\end{equation}
where $m^{\star}$ is the effective mass of the electrons.  Since the
two additive terms in the Hamiltonian must commute (one depending only
on the orbital part and the other on the spin part of the
wavefunction), we can adopt the solution of Hod {\em et
al.}~\cite{Hod2004,Hod2005b} for the transmission of electrons through
the two prototype devices (spin filter and splitter).  We adopt a
transformation $\epsilon \rightarrow \epsilon-\epsilon_{\sigma}$ which
allows us to project spin effects onto a calculation where spin was
previously neglected, noting that the spin dependent part of the
Hamiltonian under a constant field must have the two eigenvalues
$\epsilon_{\sigma}=\pm g \mu_{\mbox{\tiny B}} B$ for spin up/down
electrons.  This transformation is not limited to the specific
Hamiltonian described above: the only requirement is that $H_{ring}$
does not contain a spin dependency.  Furthermore, even for
inhomogeneous magnetic fields, where the spin-dependent term in the
Hamiltonian does not commute with $H_{ring}$, a similar transformation
in the limit of adiabatic spin dynamics can be made, where in addition
to a shift in energy one has to introduce a shift in the magnetic
flux.\cite{Richter01a}

Using the standard analytical approach of treating a 1D ring based on
a scattering matrix formalism,\cite{Gefen84} the transmission as a
function of energy, previously calculated for spinless
electrons,\cite{Hod2004,Hod2005b} need only be modified by the Zeeman
energies for up (down) electrons, which affects only the kinetic phase
angles $\phi_{k}$ in the expression:
\begin{equation}
\begin{split} 
\phi_{k}^{\uparrow \downarrow} & \equiv \phi_{k}(\epsilon'_{\uparrow
\downarrow})=\frac{\pi R}{\hbar}\sqrt{2m^{\star}\epsilon'_{\uparrow
\downarrow}} = \frac{\pi R}{\hbar}\sqrt{2m^{\star}(\epsilon \pm g
\mu_{\mbox{\tiny B}} B)}.
\end{split}
\label{eq:phik}
\end{equation}
The transmission itself is the solution of the linear scattering
problems with the final results for the two-terminal device shown in
the upper panel of Fig.~\ref{fig:sketch} given by:
\begin{equation}
T_{\uparrow \downarrow}(\epsilon) = \frac{A_{\uparrow
\downarrow}(1+\cos2\phi_{m})} {R_{\uparrow \downarrow}^{-1} +
P_{\uparrow \downarrow}\cos2\phi_{m} + Q_{\uparrow \downarrow}
\cos^{2}2\phi_{m}}
\label{eq:trans1}
\end{equation}
where $\phi_{m}=\pi\frac{\phi}{\phi_{0}}$ is the ratio between the
magnetic flux $\phi = \pi R^{2} B$ and the quantum flux $\phi_{0} =
\frac{2\pi\hbar}{e}$, and we have defined:
\begin{equation}
\begin{split}  
A_{\uparrow \downarrow} &= 16\varepsilon^{2}(1-\cos2\phi_{k}^{\uparrow
\downarrow})\\ P_{\uparrow \downarrow} &=
2(c-1)^{2}(c+1)^{2}-4(c^{2}+1)(c+1)^{2}\cos2\phi_{k}^{\uparrow
\downarrow} \\ Q_{\uparrow \downarrow} &= (c+1)^{4} \\ R_{\uparrow
\downarrow} &= (c-1)^{4} + 4c^{4} + 4 - \\
&4(c^{2}+1)(c-1)^{2}\cos2\phi_{k}^{\uparrow
\downarrow}+8c^2\cos4\phi_{k}^{\uparrow \downarrow},
\end{split}
\label{eq:par1}
\end{equation}
with $\sqrt{\varepsilon}$ the transmission amplitude into the junctions
and $c=\sqrt{1-2\varepsilon}$ the junction scattering
amplitude.\cite{Hod2004}

\begin{figure*}[htp]
\includegraphics[width=9cm]{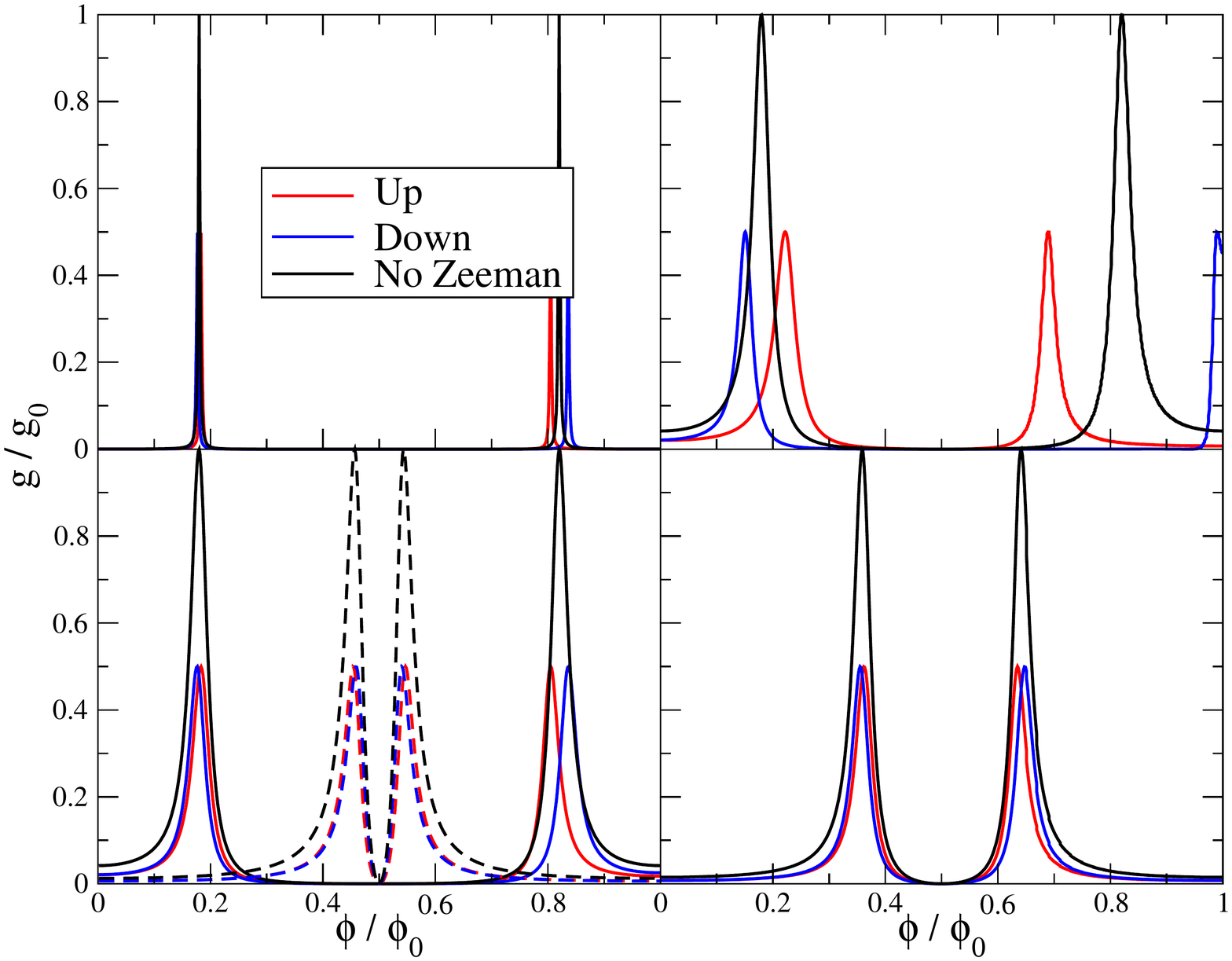}
\includegraphics[width=7.3cm]{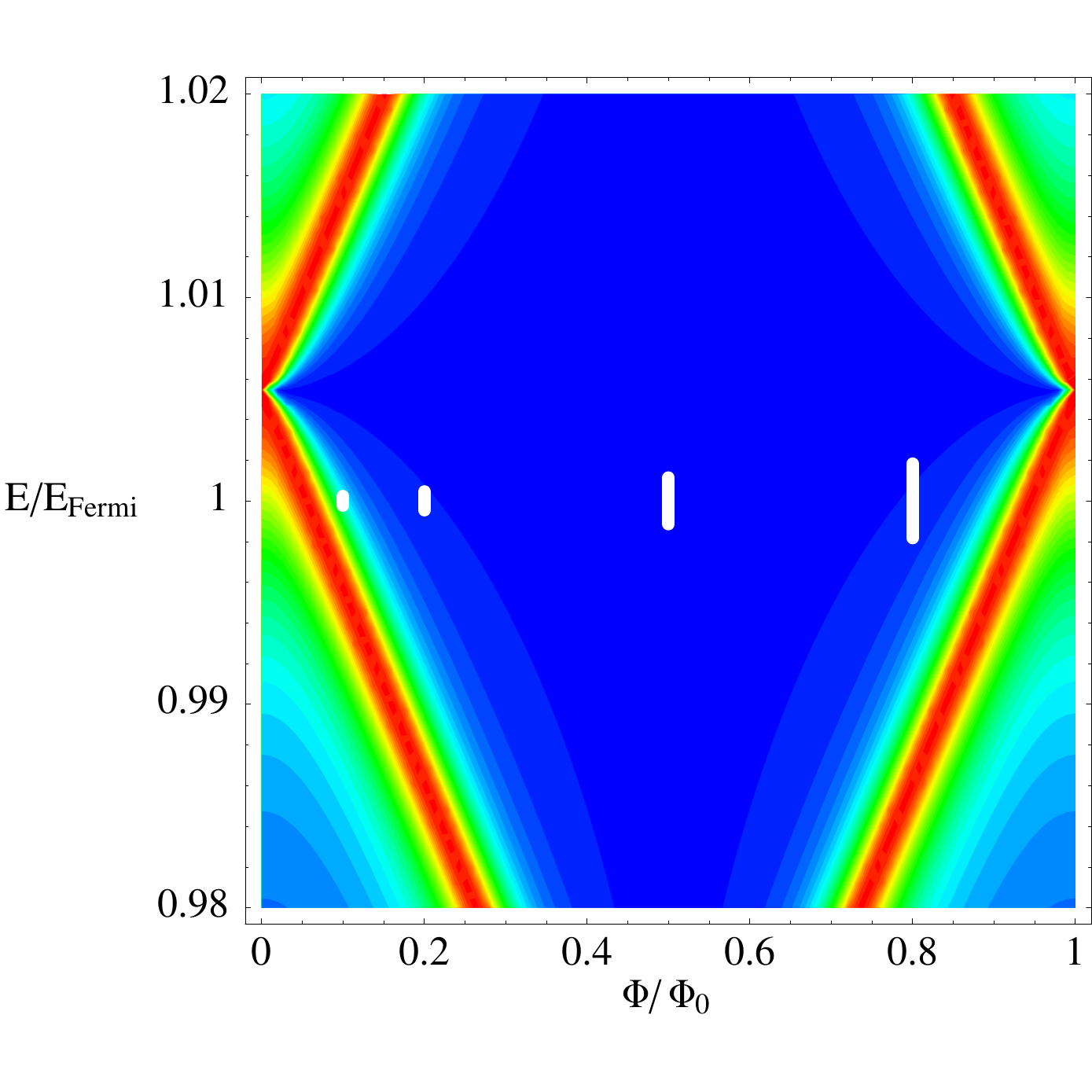}
\caption{Plots of the conduction (in units of the quantum conduction
$g_{0} = \frac{2e^2}{h}$) of a spin filter device computed within the
analytical model at zero temperature and zero bias.  Left: the
conduction as a function of the magnetic flux for different model
parameters.  The parameters used are: $R=5 \mbox{nm}$,
$\varepsilon=\frac{1}{10}$, $m^{\star}=m_e$ and $\lambda_{F}=12
\mbox{\AA}$ (solid lines) and $\lambda_{F}=4 \mbox{\AA}$ (dashed
lines) for the lower left panel, $R=5 \mbox{nm}$,
$\varepsilon=\frac{1}{100}$, $m^{\star}=m_e$ and $\lambda_{F}=12
\mbox{\AA}$ for the upper left panel, $R=10 \mbox{nm}$,
$\varepsilon=\frac{1}{10}$, $m^{\star}=m_e$ and $\lambda_{F}=12
\mbox{\AA}$ for the lower right panel, $R=5 \mbox{nm}$,
$\varepsilon=\frac{1}{10}$, $m^{\star}=10m_e$ and $\lambda_{F}=12
\mbox{\AA}$ for the upper right panel.  Right: The spinless
conductance as a function of the energy and the magnetic flux.  The
value of the conduction for the spin up/down are given by the values
at the ends of the white lines.  This figure contains the same
information as that shown in the left panels. Rainbow color scheme
used (red indicates high conductance of $\approx 1 g_{0}$).}
\label{fig:analytic spin-transmission}
\end{figure*}

A similar calculation can be made for the more cumbersome case of a
three terminal device shown in the lower panel of
Fig.~\ref{fig:sketch}.  We focus on the case where all three junctions
have identical scattering amplitudes.  The transmittance for channels
$1$ and $2$ is given by the ratio $T^{1,2}_{\uparrow \downarrow}
(\epsilon) = N^{1,2}_{\uparrow \downarrow}(\epsilon) [D^{d}_{\uparrow
\downarrow}(\epsilon)]^{-1}$.  The denominator of the transmittance
probability for both output channels is given by
\begin{widetext}
\begin{equation}
\begin{split}
D^{d}_{\uparrow \downarrow}(\epsilon) &=
\frac{1}{16}(c^2+1)(19-12c+2c^2-12c^3+19c^4) +
32c^3\cos(4\phi_{k}^{\uparrow \downarrow}) +\\
&2(c-1)^4c\{\cos[4\phi_{k}^{\uparrow \downarrow}(1-2\alpha)] +
\cos[4\phi_{k}^{\uparrow \downarrow}(1-2\beta)] +
\cos[4\phi_{k}^{\uparrow \downarrow}(1-2\gamma)]\} -\\
&8(c-1)^2c(c^2+1)\{\cos[4\phi_{k}^{\uparrow \downarrow}(\alpha-1)] +
\cos[(4\phi_{k}^{\uparrow \downarrow}\beta-1)] +
\cos[4\phi_{k}^{\uparrow \downarrow}(\gamma-1)]\} -\\
&4(c-1)^2(2-c+2c^2-c^3+2c^4)[\cos(4\phi_{k}^{\uparrow
\downarrow}\alpha) + \cos(4\phi_{k}^{\uparrow \downarrow}\beta) +
\cos(4\phi_{k}^{\uparrow \downarrow}\gamma)] +\\
&2(c-1)^4(c^2+1)\{\cos[4\phi_{k}^{\uparrow \downarrow}(\alpha-\beta)]
+ \cos[4\phi_{k}^{\uparrow \downarrow}(\alpha-\gamma)] +
\cos[4\phi_{k}^{\uparrow \downarrow}(\beta-\gamma)]\} -\\ &\frac{1}{8}
(c+1)^4 \left\{-4[1+c(c-1)]\cos(2\phi_{k}^{\uparrow \downarrow}) +
(c-1)^2\{\cos[2\phi_{k}^{\uparrow \downarrow}(1-2\alpha)] +
\cos[2\phi_{k}^{\uparrow \downarrow}(1-2\beta)]\right.\\ &\left.+
\cos[2\phi_{k}^{\uparrow \downarrow}(1-2\gamma)]\} \right\}
\cos(2\phi_m) + \frac{1}{16}(c+1)^6 \cos^2(2\phi_m).
\end{split}
\label{eq:denominator}
\end{equation}
The numerator of the transmittance probability through output channel
$1$ is given by
\begin{equation}
\begin{split}
N^{1}_{\uparrow \downarrow}(\epsilon) &=
-\frac{1}{2}\varepsilon^2\left\{-4(1+c^2)
2(c-1)^2\cos(4\phi_{k}^{\uparrow \downarrow}\alpha) +
(c+1)^2\cos(4\phi_{k}^{\uparrow \downarrow}\beta) +
2(c-1)^2\cos(4\phi_{k}^{\uparrow \downarrow}\gamma) +\right.\\&
4c\cos\left[4\phi_{k}^{\uparrow \downarrow}(\alpha+\gamma)\right] -
(c-1)^2\cos\left[4\phi_{k}^{\uparrow \downarrow}(\alpha-\gamma)\right]
-2c(c+1)\cos\left[2\phi_m-2\phi_{k}^{\uparrow \downarrow}\right] -
2(c+1)\cos\left[2\phi_m+2\phi_{k}^{\uparrow \downarrow}\right] -\\
&(c^2-1)\cos\left[2\phi_m+2(1-2\alpha)\phi_{k}^{\uparrow
    \downarrow}\right]+
(c^2-1)\cos\left[2\phi_m-2(1-2\alpha)\phi_{k}^{\uparrow
    \downarrow}\right] +
2(c+1)\cos\left[2\phi_m+2(1-2\beta)\phi_{k}^{\uparrow
    \downarrow}\right] +\\
&2c(c+1)\cos\left[2\phi_m-2(1-2\beta)\phi_{k}^{\uparrow
    \downarrow}\right] -
\left.(c^2-1)\cos\left[2\phi_m+2(1-2\gamma)\phi_{k}^{\uparrow
    \downarrow}\right]+
(c^2-1)\cos\left[2\phi_m-2(1-2\gamma)\phi_{k}^{\uparrow
    \downarrow}\right]\right\}
\end{split}
\label{eq:numerator1}
\end{equation}
and the numerator of the transmittance probability through output
channel $2$ is given by
\begin{equation}
\begin{split}
N^{2}_{\uparrow \downarrow}(\epsilon) &=
-\frac{1}{2}\varepsilon^2\left\{-4(1+c^2)
+(c+1)^2\cos(4\phi_{k}^{\uparrow \downarrow}\alpha) +
2(c-1)^2\cos(4\phi_{k}^{\uparrow \downarrow}\beta) +
2(c-1)^2\cos(4\phi_{k}^{\uparrow \downarrow}\gamma) +\right.\\
&4c\cos\left[4\phi_{k}^{\uparrow \downarrow}(\beta+\gamma)\right] -
(c-1)^2\cos\left[4\phi_{k}^{\uparrow \downarrow}(\beta-\gamma)\right]-
2c(c+1)\cos\left[2\phi_m+2\phi_{k}^{\uparrow \downarrow}\right] -
2(c+1)\cos\left[2\phi_m-2\phi_{k}^{\uparrow \downarrow}\right] +\\
&2c(c+1)\cos\left[2\phi_m+2(1-2\alpha)\phi_{k}^{\uparrow
    \downarrow}\right] +
2(c+1)\cos\left[2\phi_m-2(1-2\alpha)\phi_{k}^{\uparrow
    \downarrow}\right]
+(c^2-1)\cos\left[2\phi_m+2(1-2\beta)\phi_{k}^{\uparrow
    \downarrow}\right] -\\
&(c^2-1)\cos\left[2\phi_m-2(1-2\beta)\phi_{k}^{\uparrow
    \downarrow}\right] +
\left. (c^2-1)\cos\left[2\phi_m+2(1-2\gamma)\phi_{k}^{\uparrow
    \downarrow}\right] -
(c^2-1)\cos\left[2\phi_m-2(1-2\gamma)\phi_{k}^{\uparrow
    \downarrow}\right]\right\}.
\end{split}
\label{eq:numerator2}
\end{equation}
\end{widetext}
In the above equations $\alpha$, $\beta$ and
$\gamma=2\pi-\alpha-\beta$ are the angles between all three leads as
defined in Fig.~\ref{fig:sketch}.

The current is related to the transmittance through the Landauer
formula.\cite{Landauer1957} For the two terminal device the current is
given by
\begin{equation}
j_{\uparrow \downarrow} = \frac{e}{\pi \hbar} \int_{-\infty}^{\infty}
d\epsilon \left\{ f(\epsilon-\mu_{I}) - f(\epsilon-\mu_{O})\right\}
T_{\uparrow \downarrow}(\epsilon)
\label{eq:landauer2}
\end{equation}
where $\mu_{I,O}$ are the chemical potentials of the input/output
channels and the Fermi function is
$f(\epsilon)=\frac{1}{1+e^{\epsilon/k_{\mbox{\tiny B}}T}}$.  For the
three terminal device the current is given by
\begin{equation}
j^{1,2}_{\uparrow \downarrow} = \frac{e}{\pi \hbar}
\int_{-\infty}^{\infty} d\epsilon \left\{ f(\epsilon-\mu_{I}) -
f(\epsilon-\mu_{O_1,O_2})\right\} T^{1,2}_{\uparrow
\downarrow}(\epsilon),
\label{eq:landauer3}
\end{equation}
where $\mu_{O_1,O_2}$ is the chemical potential for output channel $1$
or $2$, respectively.  Conductance can be obtained from the current by
taking the derivative with respect to the bias voltage.

\subsection{Spin Filters}
\label{sec:spin-filter}
We now turn to discuss the application of the above results to the
construction of a spin filter device.  While for the case where the
Zeeman effect was neglected~\cite{Hod2004,Hod2005a,Hod2006a} it was
always desirable to vary the gate voltage so as to shift the
conduction peaks near zero magnetic field, in Zeeman spin devices this
heuristic is complicated by the dependence of the splitting on the
magnetic field strength.  In practice, a compromise between
realistically low magnetic fields and usable energy shifts restricts
the sought after set of parameters.  These include the kinetic phase
angle $\phi_{k}^{\uparrow \downarrow} = \frac{\pi
R}{\hbar}\sqrt{2m^{\star}\left(\epsilon \pm \frac{2 \hbar g
\mu_{\mbox{\tiny B}}}{e \pi R^{2}} \phi_{m}\right)}$ and the
transmission amplitude into the junction $\sqrt{\varepsilon}$ (cf.,
Eqs.~(\ref{eq:trans1}) and (\ref{eq:par1})).  Note that the kinetic
phase angle depends on the magnetic phase angle $\phi_{m}$.  This is
precisely where the aforementioned complication enters.

With the above convenient expressions one can calculate the
spin-dependent and spin-independent conduction for different ring
configurations and external parameters.  Typical results are shown in
Fig.~\ref{fig:analytic spin-transmission} for zero bias voltage
$\mu_{I}=\mu_{O}$ and for $T=0 \mbox{K}$.  As expected, the
spin-independent conduction is periodic with a period that is equal to
$\phi_0$.  Therefore, we plot only the first period, namely $0 \le
\phi/\phi_{0} \le 1$.  We find that within each period the
spin-independent conduction has a symmetric structure around
$\phi/\phi_{0}=1/2$, characterized by a double
peak.\cite{Hod2004,Hod2006a} This structure is caused by resonance
transmission through the energy levels of the ring.  The
spin-dependent conduction follows closely the behavior of the total
conduction.  The up/down conduction peaks are separated by the Zeeman
splittings which increase with increasing magnetic flux.

\begin{figure}[htp]
\includegraphics[width=8cm]{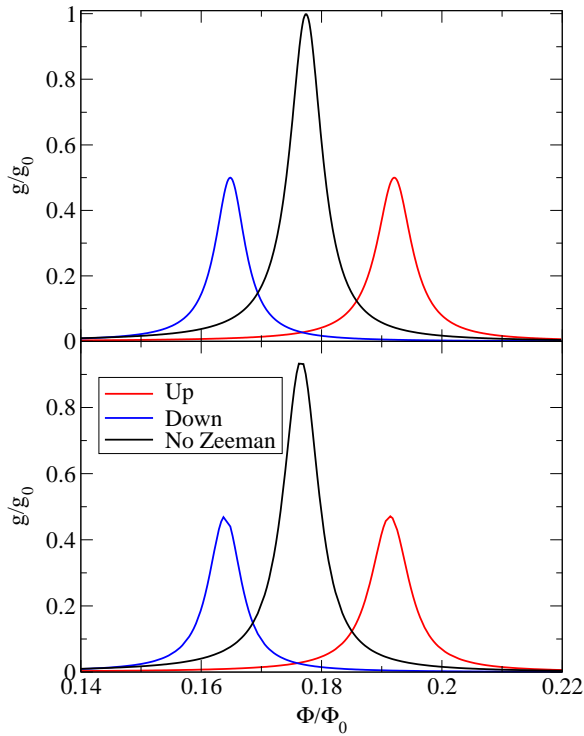}
\caption{Plots of the conduction versus magnetic flux for the
analytical model (upper panel) and for an atomistic calculation based
on a tight-binding model (lower panel). The ring radius in both cases
is $R=5 \mbox{nm}$.  Other parameters in the analytical model are
$\varepsilon=\frac{1}{50}$, $m^{\star}=4m_e$ and $\lambda_{F}=12
\mbox{\AA}$.  Tight-binding parameters are defined in the text.}
\label{fig:spinfilter}
\end{figure}

Similar to the case of spinless electrons,\cite{Hod2004} changing the
coupling strength between the leads and the ring, modeled here by
$\varepsilon$, changes the width of the peaks as shown in the left
panels of Fig.~\ref{fig:analytic spin-transmission}.  The effects of
changing the kinetic phase angle are more involved than the case
studied before.\cite{Hod2004} In the case of spinless electrons, the
kinetic phase angle depends only on the product $R
\sqrt{m^{\star}\epsilon}$.  Therefore, changing the energy of the
conduction electron by the application of a gate voltage, changing the
ring dimensions, and changing the effective mass of the conduction
electron can be mapped onto a universal curve.\cite{Hod2004,Hod2006a}
In the present study there are three independent parameters that
affect the kinetic phase angle in different ways.  For example, the
position of the conduction peaks can be shifted by adjusting the
energy of the conduction electron by the application of a gate
voltage.  This is depicted in the lower left panel of
Fig.~\ref{fig:analytic spin-transmission}.  However, the change in the
energy of the conduction electron also affects the splitting between
the conduction peaks of the up/down spins as previously discussed.
Altering the effective mass within the ring with the chemical
potential held constant, for instance by a change of composition or
inter-atomic distance in the ring atoms, modifies the kinetic phase
and can thus increase or decrease the splitting as shown in the upper
right panel of Fig.~\ref{fig:analytic spin-transmission}.  Finally, as
the ring radius is increased, the splitting for a set flux ratio
decreases since this ratio then represents a smaller field as depicted
in the lower right panel of Fig.~\ref{fig:analytic spin-transmission}.
However, for the same reason, greater magnetic flux ratios become
accessible for larger rings.

It is often more instructive to look at the results shown in the right
panel of Fig.~\ref{fig:analytic spin-transmission}, where the familiar
diamond-shaped conduction pattern within the energy/phase ratio plane
behaves more simply and predictably, in order to gain a better
intuitive understanding of the system.  Here, the two ends of the
white conduction lines represent up/down conduction where the chemical
potential and magnetic field are at the middle of such lines.  Varying
$\varepsilon$ still affects only the peak widths, modifying $R$
changes the lengths of the conduction lines, changing $m^{\star}$
scales the diagram in the energy axis without changing the length of
the conduction lines and altering the energy of the conduction
electron scales the diagram along with the lines in the same axis.
Thus, in order to construct an efficient spin filter one has to
construct this diagram and control the position of the white
conduction lines by the application of a proper magnetic field and
gate voltages to achieve a desired behavior.

Good molecular spin filters can be made when two conduction peaks with
different spins are sufficiently separated in energy and magnetic
field to allow only electrons of one spin to traverse the ring.  This
is best achieved for molecular devices with a large effective mass for
the conduction electron and small couplings between the leads and the
ring (``bad contact'').  In Fig.~\ref{fig:spinfilter} we plot the
magneto-conduction for a molecular device with relatively high
effectiveness.  We zoom on the region of the first magnetoconductance
peak.  The upper panel shows the results for a analytical model
discussed above and the lower panel shows the results of an atomistic
calculation for a molecular AB interferometer as illustrate in the
upper panel of Fig.~\ref{fig:sketch}.  As clearly can be seen,
depending on the value of the magnetic gate, one can open output
channel for spin up or down while at the same time close this channel
for the other spin.

To calculate the magneto-conductance of a molecular AB interferometer
we have used a simple tight-binding model where we assume a single
electron in a spherical $s$-level for each site on the molecular ring.
We add the proper magnetic terms to the tight-binding Hamiltonian of
the system:
\begin{equation}
H_{s} = H_{TB} + \mu_{\mbox{\tiny B}} \hat{\bf{L}} \cdot {\bf B} +
\frac{1}{2} m_{e} \mu_{\mbox{\tiny B}}^{2} B^2 R_{\perp}^{2} + g
\mu_{\mbox{\tiny B}} \mathbf{\boldsymbol{\sigma}} \cdot {\bf B},
\label{eq:Hamiltonian}
\end{equation}
where $\hat{\bf{L}}$ is the angular momentum operator, ${\bf B}$ is
the magnetic field vector, and ${\bf R}_\perp$ the projection of ${\bf
R}$ onto the plane perpendicular to ${\bf B}$.  A gauge invariant
basis is used to evaluate the tight-binding Hamiltonian matrix:
$|1s\rangle_{\alpha}^{\rm GI} = |1s\rangle_{\alpha}
e^{-\frac{ie}{\hbar}{\bf A}_{\alpha}\cdot {\bf r}}$, where
$|1s\rangle_{\alpha}$ is a $1s$ type orbital centered on site
${\alpha}$, and ${\bf A}_{\alpha}=-\frac{1}{2}({\bf R}_{\alpha} \times
{\bf B}_{\alpha})$ is the vector potential evaluated at the position
${\bf R}_{\alpha}$ of site $\alpha$.  We take the diagonal matrix
elements of $H_{TB}$ to be equal to zero (energy scale) and the
off-diagonal elements are proportional to the overlap between the
gauge invariant basis on the different electron sites, as described in
more detail in Refs.~\onlinecite{Hod2004} and \onlinecite{Hod2006a}.

The conductance is calculated using the Landauer
formalism~\cite{Landauer1957} which relates it to the scattering
transmittance probability through the system.  The transmittance is
given by $T(\epsilon)=4\mbox{Tr}[\hat{G}^\dagger(\epsilon)
\Gamma_I(\epsilon) \hat{G}(\epsilon) \Gamma_O(\epsilon)]$.  Here,
$\hat{G}(\epsilon)=[\epsilon S - H_{s}+i(\Gamma_I + \Gamma_O)]^{-1}$
is the retarded Green function, $S$ is the overlap matrix, and $\Gamma
_{I,O}$ are the imaginary parts of the self-energy ($\Sigma$) of the
input/output channels.  For the results presented in
Fig.~\ref{fig:spinfilter} we use both imaginary absorbing
potentials~\cite{Hod2004,Baer2003} and an iterative semi-infinite bulk
Green functions calculation
scheme~\cite{Hod2005a,Lopez1984,Lopez1985,Nardelli1999} to calculate
the self-energies of the leads.

Comparing the results of the simple analytical model to the results
obtained from the tight-binding model indicates that the same physical
picture emerges for the tight-binding approach.  This is expected
based on previous studies where the Zeeman effect was
neglected.\cite{Hod2004,Hod2005a,Hod2006a} The agreement between the
two approaches indicates that the diffraction pattern is insensitive
to the perturbations caused by an ionic potential and the results will
not be invalidated by a more thorough (single-particle) treatment.
The only free parameter used in the analytical theory is the
scattering amplitude $\sqrt{\varepsilon}$, which was adjusted to match
the width of the conduction peaks.  The effective mass entering the
analytical model can be calculated directly from the tight-binding
parameters. For a single $s$-level within tight-binding for a $1D$
crystal with inversion symmetry and a site distance of $a$ one can
show that the dispersion relation can be approximated
by~\cite{Merminbook}
\begin{equation}
\varepsilon_{k} \simeq E_{0}+a^{2}\gamma\left(a\right)
k^{2}.
\end{equation}
This can be compared directly with a nearly free electron of mass
$m^{\star}$ and energy $\varepsilon_{k} = E_{0} +
\frac{\hbar^{2}}{2m^{\star}}k^{2}$, allowing us to identify:
\begin{equation}
m^{\star} = \frac{\hbar^{2}}{2\gamma\left(a\right)a^{2}}.
\label{eq:effective-mass}
\end{equation}
Here $\gamma(a)$ is the resonance integral, i.e., the off-diagonal
nearest neighbor matrix element of the tight-binding Hamiltonian.


\subsection{Spin Splitter}
\label{sec:spin-splitter}
The physics of the three-terminal device shown in the lower panel of
Fig.~\ref{fig:sketch} was recently discussed for the case where the
Zeeman splitting was ignored.\cite{Hod2005b} Hod {\em et al.} showed
how cyclic molecular rings can be used as parallel magnetoresistance
logic gates (in contrast to the switching devices based on
two-terminal rings).\cite{Hod2005b} The basic idea was to couple the
cyclic molecular system to three leads creating a three terminal
device and to apply an external magnetic field.  By carefully
selecting a narrow resonance through which conductance occurred they
showed that such a setup can be used to simultaneously switch one
channel ``on'' and at the same time switch the other channel ``off''.
This was achieved by carefully adjusting the phase of the conducting
electron with diminishing amplitude on one exit channel and a large
amplitude at the other channel.  A proper combination of a gate
potential and realistically low (compared to the full AB period)
magnetic fields were used to obtain parallel logic operations such as
AND and AND+NOT.  This was demonstrated for a molecular system
composed of conjugated Benzene rings and further discussed in terms of
a single channel continuum model.

\begin{figure*}[htp]
\includegraphics[width=16cm]{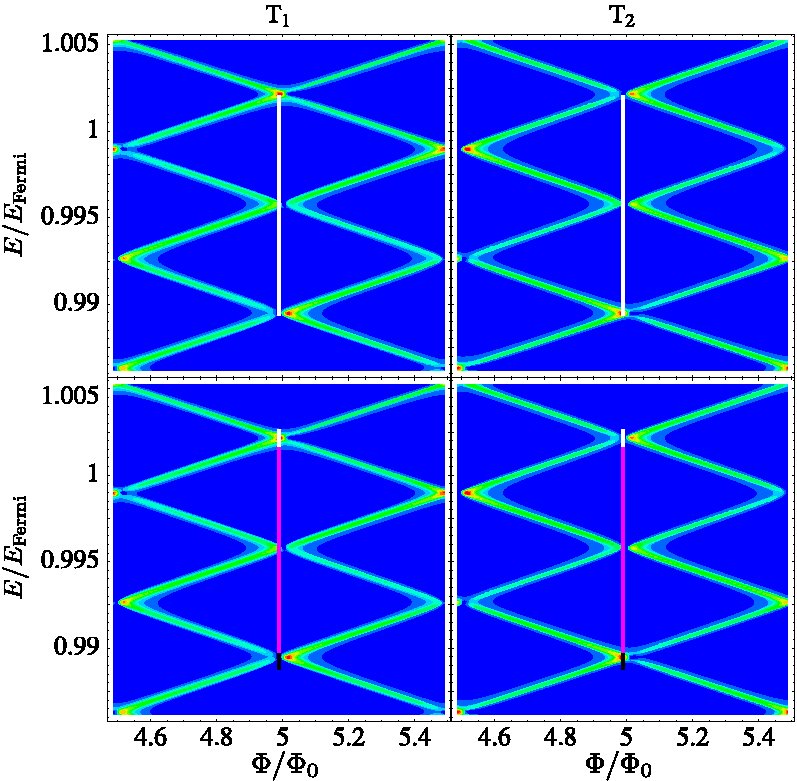}
\caption{Transmittance through a three-terminal ring device as a
function of the flux ratio (magnetic field) and the kinetic energy of
the conducting electrons.  Left and right panels correspond to output
channels $1$ and $2$, respectively.  The ring has a radius of $R=20
\mbox{nm}$, which corresponds to an AB period of about $3
\mbox{Tesla}$.  Upper panels: $m^{\star}=125 m_{e}$, zero bias, and
the two endpoints of the white line represent electrons of the two
spins.  Lower panels: $m^{\star}=10 m_{e}$, small bias of $6
\mbox{meV}$, and here, spin-independent conduction occurs all along
the purple line, while along the white (black) lines, still of length
$g\mu_{\mbox{\tiny B}}$, only up (down) electrons are transmitted, as
their conduction windows shift in energy according to their
spins. Rainbow color scheme used (red indicates high transmittance of
$\approx 1$).}
\label{fig:high-r-config}
\end{figure*}
The goal of the present study it to develop a spin splitter device by
extending the approach presented in Ref.~\onlinecite{Hod2005b} to
include the Zeeman effect.  Specifically, we will show how the phase
of the conducting electron can be tuned such that one exit channel is
turned ``on'' for up spins (``off'' for the down spins) and at the
same time the other exit channel is turned ``on'' for down spins
(``off'' for up spins).  To achieve this we need to increase the
parameter space necessary to provide means to control the efficiency
of the device.  As will become clear below, the angles between the
different channels ($\alpha$, $\beta$, and $\gamma = 2\pi - \alpha -
\beta$, cf. Fig.~\ref{fig:sketch}) will be used as control parameters
to access the many ways by which the spin splitter device can be
implemented.

We begin with a close examination of the results for the
three-terminal Zeeman spin splitter described by the simple continuum
model.  The conduction given by Eqs.~(\ref{eq:landauer3}) and
(\ref{eq:denominator})-(\ref{eq:numerator2}) for the case of zero bias
is plotted in the upper panels of Fig.~\ref{fig:high-r-config} for the
two output channels for the case of spinless electrons.  Following the
analysis of the spin filter shown in right panel of
Fig.~\ref{fig:analytic spin-transmission}, we observe the familiar
diamond-shaped conduction pattern within the energy/phase ratio plane
for each output channel.  As before, the two ends of the white
conduction lines represent up/down conduction where the chemical
potential and magnetic field are at the middle of such lines.  The
ring parameters are $R=20 \mbox{nm}$, $m^{\star}=125m_{e}$, and
$\sqrt{\epsilon}=\frac{1}{10}$.  We take $\alpha = 15.9\deg$ and
$\beta = 19.7\deg$ such that the conduction through one channel is
related to that of the other by a mirror symmetry around the field
$B_{0} \approx 5\mbox{Tesla}$: $g_{O_{1}}(B,\epsilon) \approx
g_{O_{2}}(B_{0}-B,\epsilon)$.  These angles are obtained through an
optimization procedure to achieve maximal effectiveness.

\begin{figure*}[htp]
\includegraphics[width=16cm]{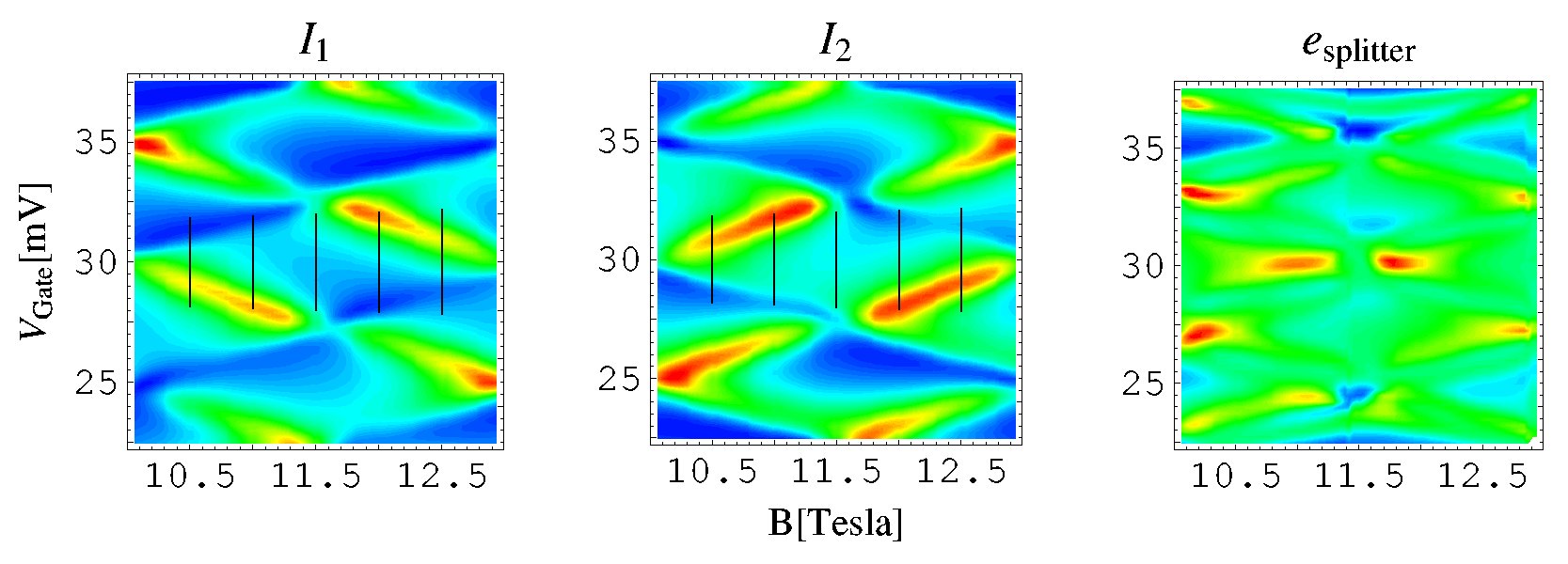}
\caption{Plots of the spinless current versus magnetic field and gate
voltage in the tight-binding approximation for the two exit channels
of a $20 \mbox{nm}$ $3$-terminal ring device (left and middle panels).
The spin dependent current is represented by the value of the current
at the ends of the black lines.  Right panel show the corresponding
current effectiveness.  In the right panel, color scale adjusted for
maximum detail, such that the red areas represent a maximum
effectiveness of approximately $0.6$ (rainbow color scheme used).  The
best polarization achieved here is $\approx 90\%$ for one channel and
$\approx 75\%$ for the other.}
\label{fig:effectiveness}
\end{figure*}

A spin splitter is obtained when one end of the Zeeman split line
conducts for one channel only (the up spin for the upper left panel of
Fig.~\ref{fig:high-r-config}) while at the same time the other end of
the Zeeman split line conducts for the other channel only (the down
spin for the upper right panel of Fig.~\ref{fig:high-r-config}).  The
structure of the conduction allows this for points where the Zeeman
separation is equal to the separation between two ring energy levels
that conduct, a very stringent requirement.  If a satisfying choice of
lead angles is always available (which appears to be the case), then
using the simple analytical model of the ring this happens when:
\begin{equation}
2 g \mu_{\mbox{\tiny B}}
B=\frac{\hbar^{2}}{2m^{\star}R^{2}}[(m+2)^{2}-(m-2)^{2}]
\label{eq:ratiorule0}
\end{equation}
Or, rewriting $B$ as $\frac{2\hbar}{e R^{2}} \frac{\phi}{\phi_{0}}$
and $\mu_{\mbox{\tiny B}}$ as $\frac{e\hbar}{2 m_{e}}$, we find that
the condition where the Zeeman separation is equal to the separation
between two ring energy levels
\begin{equation}
\frac{\phi}{\phi_{0}}=\frac{2m_{e}}{m^{\star}g} m_{\mbox {\tiny F}} =
\frac{2m_{e}}{g m^{\star}} \frac{\pi R}{a},
\label{eq:ratiorule}
\end{equation}
where, as before, $a$ is the separation between the sites on the ring.
The condition given by Eq.~(\ref{eq:ratiorule}) is a necessary but not
sufficient condition for a spin splitting device.  When the condition
is met, the angles between the input and output channels are optimized
to achieve a desired efficiency such that a mirror symmetry around the
magnetic field given by Eq.~(\ref{eq:ratiorule}) is achieved.

Similar sets of configurations with odd or half-integral flux ratio
can be found in much the same way, due to the AB splitting.  Clearly
for rings with nanometer scale dimensions, where a flux ratio of one
corresponds to thousands of Tesla, this is not a viable option unless
$m^{\star} \gg m_{e}$.  However, at tens of nanometers many
configurations are possible at fields of a few Tesla.  The main
problematic issue that remains is the fact that in order to conduct
through very low levels on the ring in this simple model, the Fermi
energy must be lowered very significantly.  If we assume that the
Fermi level is of the order of $\frac{\pi R}{a}$, one can show that in
this simple model the magnetic field required to build this device at
the Fermi energy becomes reasonable only when $R$ is of the order of
micrometers.  However, if one manages to increase the spin
$g$-factor~\cite{Majumdar98} or the effective mass in the device
$m^{\star}$ (see Eq.~(\ref{eq:effective-mass})), a nanometric device
would be feasible at mere Teslas.  This may hint that here as well a
realistic device is a matter of the choice of materials - one would in
principle need to custom-design a system in which the density of
states is high enough at low kinetic energies that only levels of such
low energies are occupied in the ground state, or where conduction
sites are far apart and weakly connected.\cite{Manoharan00,Ho03}

The problems just mentioned are no longer present if one considers
biased conduction (we assume the voltage falls symmetrically across
the junctions to avoid complication) as shown in the lower panels of
Fig.~\ref{fig:high-r-config}.  The application of a finite bias
voltage allows realistic configurations with the same structure at
various effective masses.  The example shown in the lower panels of
Fig.~\ref{fig:high-r-config} correspond to $m^{\star}=10m_{e}$ and a
bias voltage of $6 \mbox{meV}$.  The conduction window can be tuned to
contain two quite distant levels.  If levels with the transmissive
properties of the ones previously discussed are selected, the Zeeman
energy need not span the space between them.  This can be done by the
bias voltage and the Zeeman term must be no more than the level
broadening (or a few $k_{\mbox{\tiny B}}T$ if this is more) to ensure
conduction of only one spin per level.  However, the conduction of
electrons in the entire window of bias voltage should vanish in order
to make efficient spin-splitter devices, since the biased current is
the integral of the transmission over this energy range.

In an extension of the graphical method shown, the biased conduction
window could be represented by a line of length $eV_{\mbox{\tiny B}}$
(purple line), where $V_{\mbox{\tiny B}}$ is the bias voltage.  All
``spinless'' levels on this line conduct.  Centered on the line's ends
are two parallel lines of length $2 g \mu_{\mbox{\tiny B}} B$
(black/white), where only one spin level is shifted into the window -
therefore, at these ends only one spin type takes a part in
conduction, while on the rest of the lines both do.  Now, if this line
is placed like the white lines in the upper panels of
Fig.~\ref{fig:high-r-config}, with the levels at the ends in the one
spin zones, the desired result is achieved as long as the conduction
is zero for all other energies covered by the bias voltage.  This can,
however, be done at much higher energies and lower fields than those
shown in the upper panels, since the level spacing spanned by the bias
voltage can easily be orders of magnitude greater than $2 g
\mu_{\mbox{\tiny B}} B$.

In order to make a stronger argument that the spin-splitting
configurations are a physical phenomenon rather than an idiosyncrasy
of the simple one-dimensional analytical model used here, one might
proceed by reproducing them in a more detailed numerical model.  While
remaining in the independent electron picture, an obvious and flexible
choice is a ring of atoms represented by a tight-binding Hamiltonian,
for which conductance can be calculated with the methods described
above.  The wealth of parameters makes it problematic to fine-tune
exactly corresponding arrangements between the two models, and yet it
is quite a simple matter to construct in one model a ring and lead
configuration which worked well as a splitter in the other, and to try
and see how well it works at some choice of parameters which should
have analogous physical meaning.  This should be more similar to what
might be done in an actual experiment.

Since in the analytical model of a $20 \mbox{nm}$ ring we have already
located, as shown in Fig.~\ref{fig:high-r-config}, a good splitter
configuration near $10 \mbox{Tesla}$, the same configuration (with
some effective mass) would be a convincing place to look for a
splitter using the tight-binding model.  We focus on the more
realistic low effective mass regime where a bias voltage is needed and
on a finite temperature of $1 \mbox{K}$, which should still leave the
system well within the quantum regime.  In
Fig.~\ref{fig:effectiveness} we plot the current as a function of the
experimentally available parameters - the gate voltage and the
magnetic field strength.  The familiar asymmetric structure from the
previous plots is blurred, but still readily recognizable in the two
left panels, while the right panel shows the high effectiveness which
can be reached when the magnetic field, the gate voltage and the bias
voltage are all appropriately tuned.  The location of the
effectiveness peak could be easily predicted by the transmission line
method previously discussed.  Obviously, it would be practically
impossible to stumble upon such a fortunate combination of conditions
by accident, and any experiment must search for them under the
explicit guidance of a model such as the analytical model suggested
above.

\section{Coulomb Blockade Calculations}
\label{sec:coulomb-blockade-calculation}
\subsection{Charging Energy}
So far, we have ignored the energy it takes to inject multiple charges
into the small region of the ring.  This is of some concern since, as
will be discussed below, the charging energy can be very significant
here and especially when ring-leads coupling is weak, charging effects
can play a major role.  To stress this point, consider a device (ring)
that is charged with an electron of spin up.  The energy it takes to
bring another electron with a spin down can be significantly different
than in the single electron picture.  Can this additional degree of
freedom provide means to construct more efficient molecular spin
filter and spin splitter devices, or will it prevent their
realization?

\begin{figure}[htp]
\includegraphics[width=8cm]{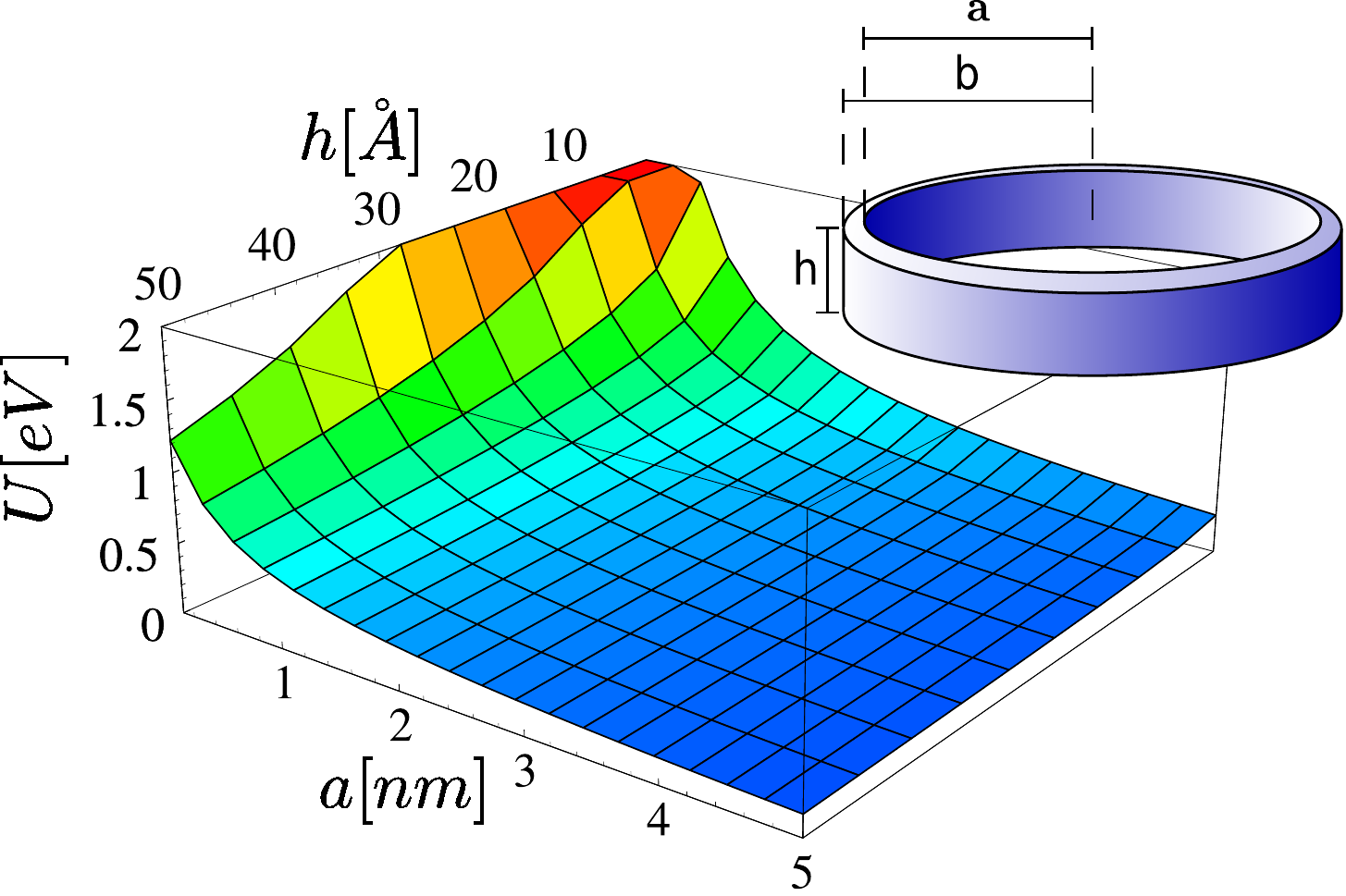}
\caption{The charging energy of a nanometric ring is plotted against
the ring's geometric parameters.  The model used is a cylinder of
radius $b$ and height $h$ with a cylindrical hole of radius $a$, where
$b$ is taken as $a+h$ to provide a square profile.  Note that the
sharp divergent peak at the origin has been cropped at $U=2 \mbox{eV}$
in order to show greater detail.}
\label{fig:The-charging-energy}
\end{figure}
 
In general, under our set of assumptions one can expect that when
charging effects are neglected, states on the ring will be grouped
into pseudo-bands with the same total occupation, having widths of
around a few level separations and separated by the order of the Fermi
energy (which is usually much greater than the level spacing).  The
addition of the charging term will shift non-neutral bands up linearly
in the charging energy with an increasing slope for charged
configurations.  One should be able to compensate for these changes by
using a gate or bias voltage, as discussed below.

One of the simplest ways of taking the charging energy into
consideration in the regime of strong charging energy and weak lead
binding is within a multi-electron master equation
approach.\cite{Beenakker91,Kinaret92,Bonet02,Hettler03,Datta2004,Elste05,Braig05,Mukamel06,Hanggi2006,Nitzan07}
The correctness of the method requires that the effects of broadening
be negligible, which is quite generally not true, but the
approximation nevertheless provides some insight into the behavior of
nanometric devices when charging has been accounted for, particularly
when a nonzero bias has been applied.  This formalism is statistical
only, and unfortunately this makes discussion of {\em perfect} devices
that send every electron to the right place more difficult than it is
in the single-electron formalism.  Alternatively, one can utilize a
perturbation treatment in the lead-device coupling, recently suggested
by K{\"{o}}nig and Gefen for an AB mesoscopic ring,\cite{Gefen02}
however, this approach is limited to off-resonant transport only.

Before we proceed to discuss the calculation of the current within the
multi-electron master equation approach, we briefly digress to the
matter of estimating the charging energy and its variation with ring
geometry.  Calculating the energy needed to (uniformly) charge a ring
or torus is an elementary electrostatics problem, though not one to
which an analytical solution exists to our knowledge.  The calculation
below provides a qualitative estimate of the magnitude of the charging
energies and the numbers should not be taken as being accurate to
within more than an order of magnitude.

The details of the calculation are as follows.  As a model we used for
simplicity a cylinder of radius $a$ and height $h$ with a cylindrical
hole of radius $b$ through its axis (see inset of
Fig.~\ref{fig:The-charging-energy} for a sketch of the model).  The
potential $\phi(r,z)$ at $r$ and $z$ can be obtained following a
standard procedure,\cite{Jackson} and is given by:
\begin{equation}
\begin{split}
\phi(r,z)&=\int_{b}^{a} r' dr' \int_{0}^{2\pi} d\varphi \int_{0}^{h}
dz' \times \\ & \frac{\rho}
{\sqrt{(r-r'\cos\phi)^{2}+r'^{2}\sin^{2}\phi+(z-z')^{2}}}
\label{eq:phi}
\end{split}
\end{equation}
The charging energy for a single electron may then be found by setting
$\rho$ to $\frac{e}{h\pi(a^{2}-b^{2})}$ inside the ring and zero
otherwise:
\begin{equation}
U_{0}=\frac{1}{2}\int\mathbf{dr}\rho\phi(\mathbf{\mathrm{r}},z)
\label{eq:U}
\end{equation} 
The charging energy for $N$ electrons is then given by
$U_{0}(N-N_{0})^{2}$ where $N_{0}$ is the neutral number of electrons.

In Fig.~\ref{fig:The-charging-energy} we plot the charging energy of a
nanometric ring as a function of the cylinder radius $a$ and height
$h$.  For typical rings considered in this work with a radius of $a=5
\mbox{nm}$, a height of $h=1 \mbox{nm}$, and $b=a+h=6 \mbox{nm}$ the
charging energy is of the order of $U_{0} < \frac{1}{2} \mbox{eV}$.
For smaller dimensions the charging energy increases considerably and
can exceed several electron volts for sub-nanometer rings.

\subsection{Calculating the Spin Current}
\label{sub:spin-current}
If one neglects spin-dependent multi-electron effects, then it is
formally straightforward to construct from the set of one-electron
Hamiltonian and spin eigenfunctions an anti-symmetric basis of
multi-electron wavefunctions:
\begin{equation}
\Psi_{n_{1}n_{2}...n_{i}}=A_{1...i}\prod_{n_{i}=1}\varphi_{n_{i}}
\label{eq:psi}
\end{equation}
Here $A_{1...i}$ is the antisymmetrization operator and the states are
identified by their (spin-dependent) level occupations $n_{i}$ ($0$ or
$1$ for fermions).  Using this anti-symmetric multi-electron
wavefunction we can uniquely and conveniently determine the matrix
elements of a general many-body operator $G$ according to the
Slater-Condon rules where only single electron integrals are taken
into account:
\begin{widetext}
\begin{subequations}
\begin{eqnarray}
& & \langle\varphi_{i}|G|\varphi_{j}\rangle=g_{ij} \\ & &
\langle\Psi_{n_{1}n_{2}...n_{M}}|G|\Psi_{n_{1}n_{2}...n_{M}}\rangle=\sum_{j}^{M}
g_{jj} n_{j} \label{eq:sumrule} \\ & &
\langle\Psi_{n_{1}n_{2}...n_{k}...n_{M}}| G
|\Psi_{n_{1}n_{2}...n'_{k}...n_{M}}\rangle = g_{kk}
\label{eq:diffrule}\\
& & \langle\Psi_{n_{1}n_{2}...n_{j}...n_{k}...n_{M}}| G
|\Psi_{n_{1}n_{2}...n'_{j}...n'_{k}...n_{M}}\rangle =
g_{jk}\delta_{n_{j}-n'_{j}-1}\delta_{n_{k}-n'_{k}+1}
\label{eq:crossrule}\\
& & \langle\Psi_{n_{1}n_{2}...n_{j}...n_{k}...n_{l}...n_{M}}| G
|\Psi_{n_{1}n_{2}...n'_{j}...n'_{k}...n'_{l}...n_{M}}\rangle=0,
\label{eq:zerorule}
\end{eqnarray}
\end{subequations}
\end{widetext}
where $M$ is the number of single-electron levels taken into account,
and $n'_{i}=|1-n_{i}|$.  Multi-electron effects will be considered
only in the form of charging energy.  Since these values will be used
in a rate-process calculation rather than a full quantum formulation,
constructing the multi-electron states themselves is actually
redundant, and Eqs.~(\ref{eq:sumrule})-(\ref{eq:zerorule}) along with
the single particle data will provide all the necessary information.

In order to perform a master equation based estimation of the current,
the transfer rates between different multi-electron states must first
be
calculated.\cite{Beenakker91,Kinaret92,Bonet02,Hettler03,Datta2004,Elste05,Braig05,Mukamel06,Hanggi2006,Nitzan07}
We will assume that only levels near the Fermi energy will take part
in conduction.  This implies that the $N$ levels closest to the Fermi
energy will be used to construct the $2^{N}$ multi-electron states
themselves using Eq.~(\ref{eq:sumrule}), with the additional charging
term $U_{0}(N-N_{0})^{2}$.

The transfer rate through lead $\ell$ between two multi-electron
states is given by:
\begin{equation}
R_{\ell,\alpha\rightarrow\beta} =
\frac{\Gamma_{\ell,\alpha\beta}}{\hbar}Q_{\alpha\beta},
\label{eq:rate1}
\end{equation}
We label multi-electronic states by the Greek indices $\alpha \equiv
\{n_{1}^{(\alpha)}, n_{2}^{(\alpha)} \cdots n_{N}^{(\alpha)}\}$ and
$\beta \equiv \{n_{1}^{(\beta)}, n_{2}^{(\beta)} \cdots
n_{N}^{(\beta)}\}$.  Single electronic levels are labeled by the
indices $i$ and $j$.  We also define the total transfer rate summed
over all leads:
\begin{equation}
R_{\alpha\rightarrow\beta} =
\sum_{\ell}R_{\ell,\alpha\rightarrow\beta}.
\label{eq:rate2}
\end{equation}
In the above equations, $\Gamma_{\ell,\alpha\beta}$ is related to the
the imaginary part of the self-energy in the single electron picture.
To lowest order, $\Gamma_{\ell,\alpha\beta}=\gamma_{\ell,ii}$ if the
two multi-electronic states differ only by the occupation of level
$i$, $\Gamma_{\ell,\alpha\beta}=\gamma_{\ell,ij}$ if they differ only
by $n_{i}$ and $n_{j}$ and $n_{i}-n_{j}=1$.  Otherwise
$\Gamma_{\ell,\alpha\beta}=0$.  This follows from the Slater-Condon
rules (cf., Eqs.~(\ref{eq:diffrule}) and (\ref{eq:crossrule}).
$\gamma_{\ell,ij}$ is the matrix element of the imaginary part of the
self-energy.  $Q_{\alpha\beta}$ in Eq.~(\ref{eq:rate1}) is related to
the Fermi-Dirac function, $f(\epsilon)$:
\begin{equation}
Q_{\alpha\beta}=
\begin{cases}
f(\epsilon_{\alpha}-\epsilon_{\beta}-\mu_{\ell}) &
N_{\alpha}>N_{\beta},\\
1-f(\epsilon_{\alpha}-\epsilon_{\beta}-\mu_{\ell}) & N_{\alpha} <
N_{\beta},\\ 1 & N_{\alpha} = N_{\beta},
\end{cases}
\label{eq:Q}
\end{equation}
where $N_{\alpha}=\sum_{i} n_{i}^{(\alpha)}$ is the number of
electrons in state $\alpha$.

\begin{figure*}[htp]
\includegraphics[width=7cm]{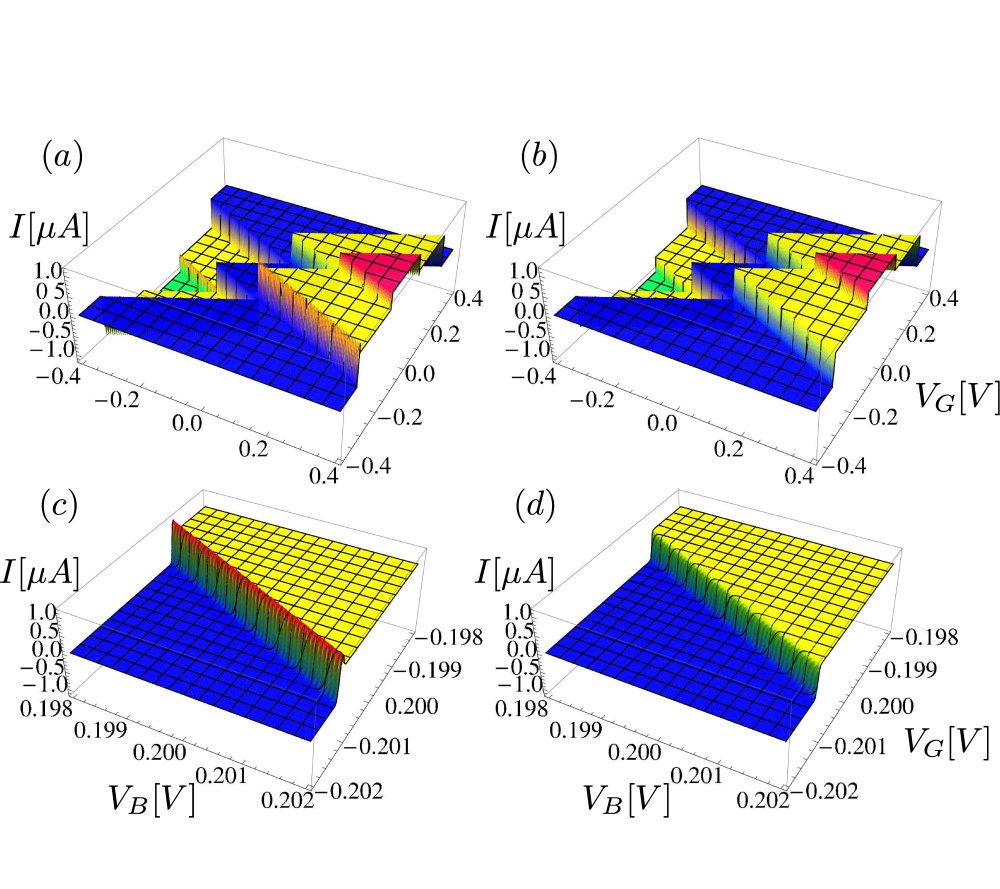}
\includegraphics[width=8cm]{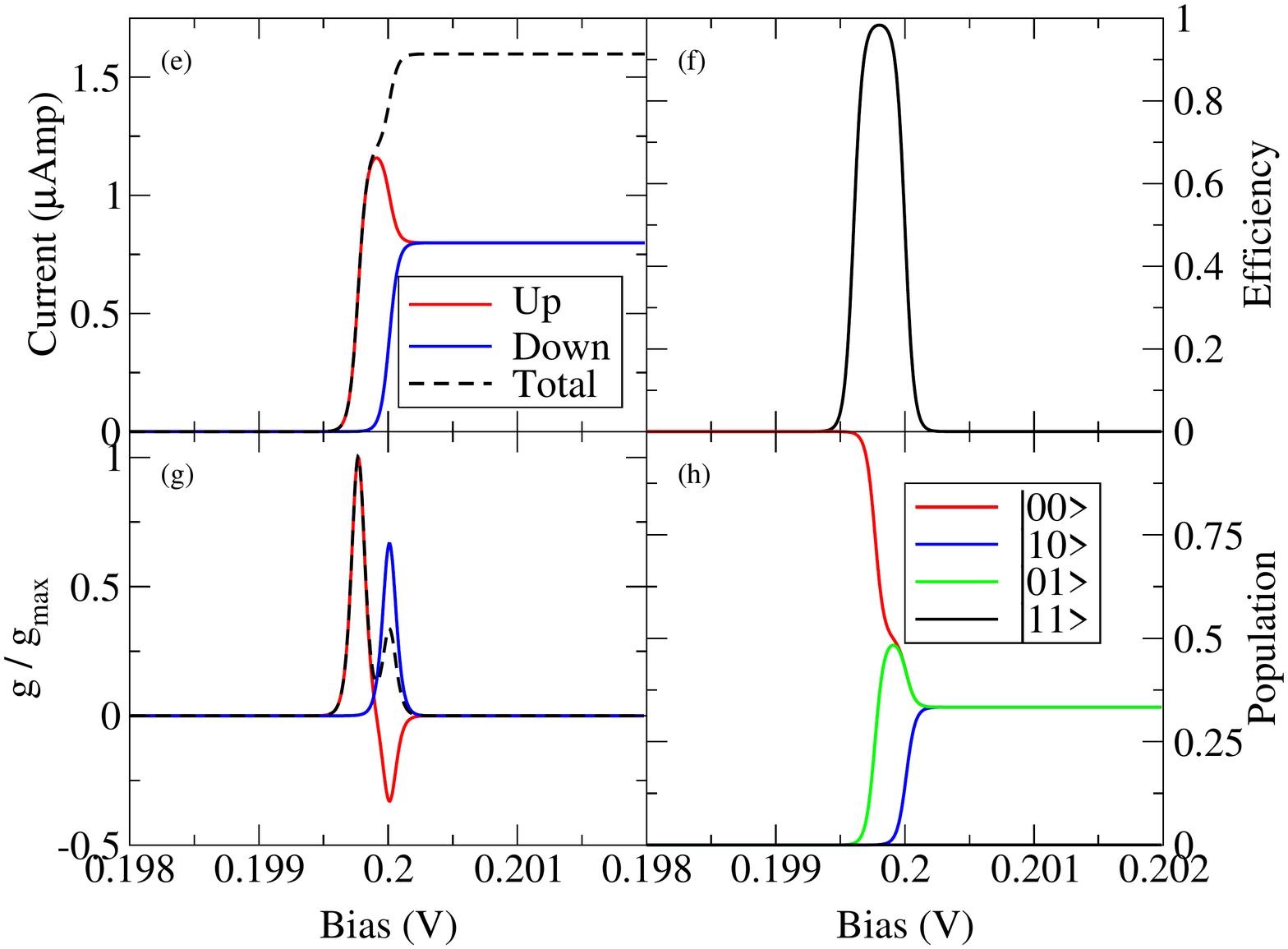}
\caption{Panels (a)-(d): surface plots of the current versus gate and
bias voltages.  (a),(c) and (b),(d) panels correspond to up-spin and
down-spin currents, respectively.  The lower panels zoom into a region
of the voltage plane in which a realization of an up spin-filter is
possible.  Panels (e)-(h): a cut through constant gate voltage plane
of the current (e), the effectiveness (f), the conduction (g), and the
state population (h) as a function of the bias voltage.  In all panels
the ring is composed of $40$ single electron sites with metallic leads
at opposite ends, and a magnetic field of $10 \mbox{Tesla}$ is
applied.  The charging energy is $\frac{1}{10} \mbox{eV}$ and
$T=\frac{1}{5} \mbox{K}$.
\label{fig:ME-filter}}
\end{figure*}

With the rates known, the linear master equation system can be written
as the condition for steady-state:
\begin{equation}
\sum_{\beta}R_{\alpha\rightarrow\beta}P_{\alpha} -
\sum_{\beta}R_{\beta\rightarrow\alpha}P_{\beta}=0,
\label{eq:master}
\end{equation}
where $P_{\alpha}$ is the probability that the system is in a
multi-electron state $\alpha$.  Once the steady-state occupation
probabilities have been solved for the current can be expressed
as\cite{Beenakker91,Kinaret92,Bonet02,Hettler03,Datta2004,Elste05,Braig05,Mukamel06,Hanggi2006,Nitzan07}:
\begin{equation}
I_{\ell} = -e \sum_{\alpha\beta} R_{\ell,\alpha\rightarrow\beta}
P_{\alpha} s_{\alpha\beta},
\label{eq:current1}
\end{equation}
where
\begin{equation}
s_{\alpha\beta} = \begin{cases} +1 & N_{\alpha}<N_{\beta},\\ -1 &
N_{\alpha}>N_{\beta},\\ 0 & N_{\alpha}=N_{\beta}.
\end{cases}
\label{eq:s1}
\end{equation}
Intuitively, this expression states that current flows out of lead
$\ell$ whenever an electron flows from it into the device, with the
inverse also true.  Following a similar line of physical reasoning
leads to an expression for spin-polarized current: up or down current
flows out of lead $\ell$ whenever an up or down electron flows from it
into the device.  Assuming no coupling between levels with different
spin, the spin-dependent current is given by:
\begin{equation}
I_{\ell,\uparrow(\downarrow)} = -e\sum_{\alpha\beta}
R_{\ell,\alpha\rightarrow\beta} P_{\alpha}
s_{\uparrow(\downarrow)\alpha\beta},
\label{eq:current2}
\end{equation}
and
\begin{equation}
s_{\uparrow(\downarrow)\alpha\beta} = \begin{cases} +1(0) &
S_{\alpha}<S_{\beta}\wedge N_{\alpha}<N_{\beta},\\ 0(+1) &
S_{\alpha}>S_{\beta}\wedge N_{\alpha}<N_{\beta},\\ 0(-1) &
S_{\alpha}<S_{\beta}\wedge N_{\alpha}>N_{\beta},\\ -1(0) &
S_{\alpha}>S_{\beta}\wedge N_{\alpha}>N_{\beta}.
\end{cases}
\label{eq:s2}
\end{equation}
Here, $S_{\alpha}=\sum_{i} s_{i}^{(\alpha)}$ where $s_{i}^{(\alpha)} =
\pm 1$ for spin up or down, respectively.

Peaks in the differential conduction as a function of the bias voltage
$V_{\mbox{\tiny B}}$ can be expected whenever there exists a
difference in energy between two states differing in their number of
electrons by one, which is occupied in one lead but not the other,
i.e., when the conduction window grows to contain a spectral line.
This is why levels spaced more than about $eV_{\mbox{\tiny
B}}+k_{\mbox{\tiny B}}T$ from the Fermi level should not take part in
conduction within this formalism: transfer through them involves
electrons or holes (referring here to level vacancies) not present in
the leads.

\subsection{Spin Filter}
\label{sec:spin-filter-me}
A filtering device remains straightforward in this formalism, and the
previous discussion in subsection~\ref{sec:spin-filter} pertains to it
as well.  Several differences are nevertheless evident.  In the single
electron formalism a device will conduct at zero bias when the
chemical potential coincides with an energy level on the device.  In
the master equation formalism this still happens when the charging
energy is zero.  Under such conditions, the difference in energy
between the neutral state and the first charged state always equals
the Fermi energy, and this difference is obviously contained in the
zero-bias conduction window.

In the presence of charging, the picture is somewhat more involved.
The same states discussed above are now shifted differently by
charging since they have a different number of electrons.  For the
sake of clarity, we will first consider the simplest case where only
$2$ levels and thus $4$ states are included.  We assume that the
system can have $N_{0}-1$ electrons, $N_{0}$ electrons, or $N_{0}+1$
electrons with energies $U_{0}$, $\epsilon \pm g \mu_{\mbox{\tiny B}}
B$, and $2\epsilon + U_{0}$, respectively ($\epsilon$ is the single
electron level energy). Furthermore, $N_{0}$ is the neutral occupation
and we assume it is such that one of the aforementioned levels is
occupied when the systems is neutral.  The value of these energies was
calculated by taking the single electron level $\epsilon$ and adding
charging terms ($U_{0}$ for both $N_{0}-1$ and $N_{0}+1$) and magnetic
field splittings (for the $N_{0}$ state).

When the chemical potential in one of the leads equals the energy
difference between multi-electron states a peak will appear in the
differential conductance.  For the case described here when the bias
voltage falls symmetrically on both junctions, conduction peaks will
occur when the energy difference between two multi-electron states of
different occupations is equal to the value of the lead Fermi energy
combined with half the applied bias voltage.  Taking the Fermi energy
as that of the lower of the two single electron conduction levels, the
conduction peaks are expected to appear at $eV_{\mbox{\tiny B}} = 2
U_{0}$ and $eV_{\mbox{\tiny B}} = 2 (U_{0} + 2 g \mu_{\mbox{\tiny B}}
B)$ (these peaks are doubly degenerate).

Typical behavior of a spin filter device is illustrated in
Fig.~\ref{fig:ME-filter}, where $U_{0}=\frac{1}{10} \mbox{eV}$,
$\epsilon=1 \mbox{eV}$, and $T = \frac{1}{5} \mbox{K}$.  On the left,
surface plots of the current as a function of the gate and bias
voltages are shown (panels (a)-(d)). Upper panels show the current of
spin up (left) and spin down (right) electrons for a wide range of
gate and bias voltages.  Note how current steps (or conduction peaks)
occur whenever the conduction window cuts across a difference between
state energies as discussed above.  For instance, with no gating
transitions with $N \rightarrow N \pm 1$ occur simultaneously at
$eV_{\mbox{\tiny B}}=2U_0$, and with a gate voltage of $\pm U_0$ the
first step occurs at zero bias.  On the scale of the plots shown in
panels (a) and (b), the currents for the two different spins are
almost indistinguishable.  The difference in current between the two
polarization can be seen in panels (c) and (d), where we zoom in on a
specific area of the line described by $eV_{\mbox{\tiny
G}}=U_0-eV_{\mbox{\tiny B}}/2$, where $V_{\mbox{\tiny G}}$ is the gate
voltage.  Here, one spin starts to flow at slightly lower bias
voltages, the current rising with the voltage to a high peak and then
falling back down as the other spin begins to flow as well.  A plane
cut through this surface is shown in panel (e), where it is clearly
demonstrated how by tuning the voltages a spin filter can still be
constructed when charging is taken into account.  A symmetric filter
for the opposite spin type is found when inverting the gate voltage.
As can be seen in the figure, the first conduction peak occurs at
$eV_{\mbox{\tiny B}} = 2(U_{0}-2g\mu_{\mbox{\tiny B}}B)$ and the
second occurs at $eV_{\mbox{\tiny B}} = 2U_{0}$.  As a result of the
broken symmetry, these peaks are non-degenerate and correspond to
transition with $N \rightarrow N-1$.  The other two peaks
corresponding to $N \rightarrow N+1$ will occur at a significantly
higher bias voltage.  This is in contrast to the simple example where
only $4$ states were considered, as describe above, where the current
peaks are degenerate and appear simultaneously.

One of the interesting features shown in Fig.~\ref{fig:ME-filter} is
the negative differential {\em spin} conductance of the spin up
electron (panel (g)).  This is explained by the sudden drop in the
population of the $|01>$ state (corresponding to a conduction electron
of spin up) as the change in chemical potentials begins to allow the
population of state $|10>$ (corresponding to a conduction electron of
spin down).  This population switching is reminiscent of the
nonmonotonic change of occupation in two electrostatically coupled
single-level quantum dots.\cite{Gefen05,Oreg05} The state populations
are shown in panel (h) of Fig.~\ref{fig:ME-filter}.  The drop in the
spin up state population is correlated with the occurrence of negative
differential {\em spin} conductance (the conduction is normalized to
the maximal value~\cite{Beenakker91}).  This can be explained in the
following way: the current for each spin is determined by the product
of the probability that the system is in state $\alpha$ and the rate
of transitions between state $\alpha$ and state $|00>$ (which is the
state with no conducting electrons), where $\alpha$ is $|10>$ for spin
up conducting electron and $|01>$ for spin down conducting electron.
We therefore expect that at chemical potentials where the relevant
Fermi functions and hence the rates are nearly constant, the current
will be approximately proportional to the population.

\subsection{Spin Splitter}
The discussion of the spin splitter within the multi-electron master
equation is more involved than the spin filter and requires at least
$4$ levels ($16$ states).  We assume that at $N_{0}$ (the neutral
occupation) two of the four conduction levels are occupied.  Having
previously established a way in which single electron levels suitable
for spin splitting can be found on a ring, we focus on the effects of
charging.  First, and for the moment neglecting charging energies, we
assume that we have found two levels with energies $\epsilon \pm
\Delta$ that are spin degenerate without Zeeman splitting, with the
following properties: they should be adjacent (other than perhaps for
non-transmissive levels, which will not qualitatively affect our
results), and should both be coupled symmetrically to the input lead,
while each is coupled to a single different output lead.  Other levels
should be at least $2\Delta$ away in energy.  All couplings are taken
to be the same and diagonal couplings are ignored.  Accounting for
Zeeman splitting will simply shift each of these levels in energy by
$\pm g \mu_{\mbox{\tiny B}} B$ without modifying the coupling.  As
before for the single electron picture, the coupling between levels
with different spins is neglected.  An example of such a case is
discussed above in Sec.~\ref{sec:spin-splitter} where $\Delta$ is
given in terms of the ring level spacing (c.f.,
Eq.~\ref{eq:ratiorule0} and Fig.~\ref{fig:high-r-config}).

\begin{figure}[htp]
\includegraphics[width=8cm]{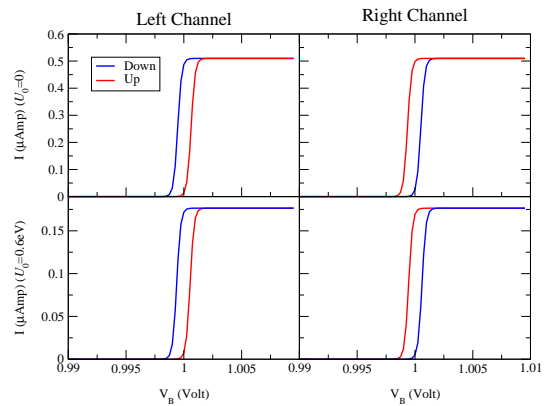}
\caption{A spin-splitter made from two levels $1 \mbox{eV}$ apart
where $\Delta = \frac{1}{2} \mbox{eV}$, each of which is further split
by the Zeeman energy at $10T$.  The upper diagrams show the
conductance through the two output channels when the charging energy
is ignored and the lower diagrams shows the conductance of the two
output channels at a charging energy of $U_{0} = 0.6 \mbox{eV}$.  We
have adjusted the gate potential so that in both cases the current
rises at $V_{\mbox{\tiny B}} \approx 1 \mbox{Volt}$ and at this value
only one spin conducts through each lead.}
\label{fig:ME-Splitter}
\end{figure}

Setting the chemical potential of the input lead to $\mu_{I} =
\epsilon + \frac{e V_{\mbox{\tiny B}}}{2}$ and both output chemical
potentials to $\mu_{O} = \epsilon - \frac{e V_{\mbox{\tiny B}}}{2}$,
it should now be clear under both formalisms that without charging
energy a perfect spin splitter exists at $eV_{\mbox{\tiny B}} =
2\Delta - 2g\mu_{\mbox{\tiny B}}B$, as shown in the upper panels of
Fig.~\ref{fig:ME-Splitter}.  In the single electron picture this
happens as two levels with opposite spins and different lead bindings
enter the conduction window.  In the multi-electron master equation
this happens when the difference in energy between $N=N_{0}$ states
and $N=N_{0} \pm 1$ states enters the conduction window.  These two
conditions are equivalent for $U_{0}=0$.

Including charging effects will cause the $N = N_{0} \pm 1$ states to
move up in energy by $U_{0}$.  This has the effect of repositioning
the conductance peaks at $eV_{\mbox{\tiny B}} = 2\Delta \pm
2g\mu_{\mbox{\tiny B}}B \pm U_{0}$.  If $U_{0}$ is of the order of the
Zeeman term then the charging energy will interfere with the structure
that allowed our spin splitter for $U_{0}=0$, since the energy
difference between states that differ by one electron of both spin up
and spin down can become similar.  On the other hand, if the charging
energy is increased further, the transitions $N=N_{0} \leftrightarrow
N+1$ and $N=N_{0} \leftrightarrow N-1$ become distinguishable in
energy.  One way to take advantage of this is to apply a gate
potential in order to shift the state energies by $-U_{0}$ so that
only the $N=N_{0} \leftrightarrow N+1$ transitions are in the
conduction window (new transitions may appear at similar energies if
other levels are too close, but since level spacing for a ring
generally increases with energy we can neglect such contributions).
Once again at $eV_{\mbox{\tiny B}} = 2\Delta - 2g\mu_{\mbox{\tiny
B}}B$ perfect splitting occurs, though at a smaller total current,
since fewer state transitions are involved.  This is apparent in the
lower panels of Fig.~\ref{fig:ME-Splitter}.

\section{Conclusions and Directions for Further Study}
\label{sec:conclusions-directions-for-further}
We have investigated several ways in which rings or ring-like
structures with a radius of the order of nanometers, coupled to
metallic leads, might be used to construct simple spin-sensitive
devices.  We have focused only on the Zeeman splitting to
differentiate between spins, since it is always present while other
effects commonly utilized in microscopic structures, such as the
Rashba and Dresselhaus effects, are generally absent in
molecular/nanometric structures.  We believe the niche this work
occupies in the search for nanospintronic devices is yet unexplored:
very small coherent structures and weak spin-dependent effects have
not received much attention, despite their formal simplicity and
significance.  What we have shown here is that although building
devices under the burden of such limitations is difficult, it is
possible.  Considering the scientific and technological benefits of
such devices we believe it will also be worthwhile.

The basic calculations we have performed are enough to point one in
the right direction as to the desired properties of molecule-sized
Aharonov-Bohm spin devices and the conditions which their desired
operation might be observed.  A similar methodology can be applied to
more complex devices or sets of devices.  The actual devices
specifically discussed here were two of the most basic - a spin-filter
and a spin-splitter or Stern-Gerlach device.  However, the conclusions
drawn and principles laid out may easily be extended to many
interesting systems, from quantum gates (since in theory we can use
interference devices to perform general unitary transformations
between input and output gates) to molecular memory (since an electron
trapped on the device will modify its electrical properties and thus
may be detectable at a later time).

Several complementary methods were utilized during the course of this
investigation: first, a simple single-electron, analytical model in
which the parameter space can easily be explored, and thus basic
intuition about the system can be gained.  Second, a tight-binding
nonequilibrium Green's function treatment, which incorporates a more
realistic physical structure that can be compared directly with
experimental data, but still assume independent electrons, was applied
to similar systems studied within the simple analytical model.
Finally, a multi-electron master equation approach that can be used to
examine many-particle effects was grafted onto the tight-binding
results.  Here, we choose to focus on electric charging, since its
effects are energetically dominant.

Our analysis of the filter, meant to be a test-case, was based on the
study of the spinless case where it was possible to create very narrow
conduction peaks near zero magnetic field by combining weak
device/leads couplings and a gate voltage to shift the conduction
resonances to $B=0$.\cite{Hod2004,Hod2005a,Hod2005b,Hod2006a} Since
the Zeeman splitting depends linearly on the magnetic field, a spin
filter always requires a finite magnetic field, at least high enough
to separate the spin conduction peaks in energy by more than
$k_{\mbox{\tiny B}}T$.  Control over the position of the
spin-dependent conduction peaks and their widths can be achieved by
carefully adjusting the kinetic phase $\phi_{k}^{\uparrow \downarrow}
= \frac{\pi R}{\hbar}\sqrt{2m^{\star}\left(\epsilon \pm \frac{2 \hbar
g \mu_{\mbox{\tiny B}}}{e \pi R^{2}} \phi_{m}\right)}$ and the
coupling between the device and the leads, respectively.  Unlike the
spinless case where $\phi_{k}=\frac{\pi
R}{\hbar}\sqrt{2m^{\star}\epsilon}$, the kinetic phase is now a
function of the magnetic flux itself, and thus, the conduction is not
a simple periodic function of the parameters.  High efficiency spin
filters are constructed at the highest magnetic field possible, where
spin-dependent effects are strongest, and flexibility is gained by
selecting materials or structures with high effective mass (or large
ring size, which is not desirable).  Charging effects do not
drastically modify this picture since spin selectivity depends only
the Zeeman term and the application of a gate voltage can compensate
for the charging energy itself.  However, charging leads to a breaking
of symmetries and as a result to negative differential
spin-conduction.

The physics of the spin splitter device is similar to that described
above with the added complication that differentiated control of the
spin-dependent wavefunction at the different leads is required.
Within the single electron picture, a spin splitter device may operate
when two spin-degenerate levels exist such that one transmits through
one lead only and the other through the other lead only.  Furthermore,
the level separation should equal exactly the Zeeman splitting.
Alternatively, the level separation can be compensated for by the
application of a bias voltage, as long as there are no transmissive
levels between the two spin-degenerate levels.  Here, parameter space
includes the ring size, effective mass, device-leads couplings, and
the angles between the output leads.  Increasing the effective mass,
as before, reduces the level spacing on the ring, thereby enabling the
construction of spin splitters at lower magnetic fields and with
smaller dimensions. Charging breaks the symmetry and can reduce the
overall current through the device.  However, charging does not
abolish the general picture and the application of a gate potential
can be used to overcome most of its effects.

For both prototype devices we have shown how the limits of
lead-coupling, system geometry and temperature at which one might
expect to see the desired effects can easily be estimated, as well as
several ways in which one can implement specific behaviors by
systematically finding parameters at which they occur, either exactly
or approximately.  We have also shown that even within the parameters
space of a device formed by a single ring with two or three leads,
nontrivial behaviors with useful properties take place.  In at least
two instances, we have made a case that our findings are physical
rather than a mathematical peculiarity of some model by reproducing
them under different assumptions and formalisms.  We have found an
interesting example of negative differential conductance for a spin
polarized current with a simple explanation.  Finally, we have
described what is, to our knowledge, the smallest Stern-Gerlach
apparatus ever reasonably conceived of, and one which is fully
switchable in directionality at constant magnetic field by the
application of an external electric field.

Despite their simplicity, the calculations reported here provide
several novel predictions that seem to be model independent.
Nevertheless, more refined models are an important set of directions
to continue along.  In particular, drawing from the study of crystals
and mesoscopic systems, we see ways of enhancing the effect of Zeeman
splitting which require a more elaborate electronic structure
description.  One such is through the effects of spin-orbit coupling
and the local spin density exchange-correlation energy, both of which
have been used to explain the giant spin $g$-factor enhancement that
has been observed in mesoscopic structures under certain
conditions.\cite{Majumdar98} It is quite possible that the same effect
can be recreated with discrete levels, although this probably requires
that only a small number of levels be occupied (corresponding once
again to very low electron densities).

While we have studied only two- and three-terminal devices, with only
one injective terminal, it is reasonable that four terminal devices
will also be of interest as coherent quantum gates.  One of the
reasons we have found it worthwhile to draw attention to the
importance of many-particle effects is our hope that in time-dependent
calculations they may be used to create sequential logical behavior
without sacrificing coherence - for instance, an electron may only be
able to enter the device through one lead, but when it enters it opens
up another lead, and the inflow of another electron causes both to be
discharged through a third lead - a sort of sequential AND gate, two
of which could form a true AND gate.  Of course, to consider useful
computation it is necessary to model entire networks of such gates,
where input electrons enter on one set of leads, propagate throughout
the network and leave it on another set of leads.  Such a network
presents new delocalized challenges unless devices are somehow coupled
in such a way that no inter-device interference takes place.  However,
even if the devices within the network are all "imperfect" devices
that only approximate logic gates, as long as every device has
independently adjustable parameters (like a gate voltage) such a
network forms a fascinating basis for a model of a quantum neural
networks.  These and related issues are still open for future study.

\section{Acknowledgments}
\label{sec:acknowledgments}
We would like to thank Roi Baer, Yuval Gefen, Andrew Millis, Yuval
Oreg, Yoram Selzer, and Amir Yacoby for discussions and suggestions.
This work was supported by the EU under the program SA-NANO (grant
number STRP 013698).

\end{document}